\documentclass[structabstract]{aa}  
 \usepackage{longtable}
\usepackage{epsfig}
\usepackage{graphicx}
\usepackage{txfonts}
\usepackage{amssymb}
\usepackage{natbib}                               
\bibpunct{(}{)}{;}{a}{}{,}

\newcommand{\be}{\begin{equation}}
\newcommand{\ee}{\end{equation}}
\newcommand{\bea}{\begin{eqnarray}}
\newcommand{\eea}{\end{eqnarray}}
\def\fcol{f_{\rm c}}

\begin{document}

 \title{X-ray bursting neutron star  atmosphere models: spectra and color corrections}

\author{
V.~Suleimanov\inst{1,2} 
\and
J.~Poutanen\inst{3} 
\and
K.~Werner\inst{1}}

\offprints{V.~Suleimanov}

\institute{
Institute for Astronomy and Astrophysics, Kepler Center for Astro and
Particle Physics,
 Eberhard Karls University, Sand 1,
 72076 T\"ubingen, Germany \\
  \email{suleimanov@astro.uni-tuebingen.de,werner@astro.uni-tuebingen.de}
\and
Kazan Federal University, Kremlevskaja str., 18, Kazan 420008, Russia
        \and
            Astronomy Division, Department of Physics,
            PO Box 3000, FIN-90014 University of Oulu, Finland\\
            \email{juri.poutanen@oulu.fi}
}

\date{Received xxx / Accepted xxx}

   \authorrunning{Suleimanov et al.}
   \titlerunning{Model atmospheres of X-ray bursting NSs}

\abstract
{}
{
X-ray bursting neutron stars in low mass X-ray binaries constitute an 
  appropriate source class to constrain masses and radii of neutron stars, but a
 sufficiently extended set of corresponding model atmospheres is necessary for
 these investigations.
}
{
We computed such a set of model atmospheres and emergent spectra in a
plane-parallel, hydrostatic, and LTE approximation with
  Compton scattering taken into account.
 }
{The models were calculated for six different
chemical compositions: pure hydrogen and pure helium atmospheres, and atmospheres
with solar mix of hydrogen and helium, and various heavy element abundances
$Z$ = 1, 0.3, 0.1, and 0.01 $Z_\odot$. For each chemical composition the models are
computed for  three values of surface gravity,
$\log g$ =14.0, 14.3, and 14.6, and for 20 values of the luminosity in units of the Eddington luminosity,
$L/L_{\rm Edd}$,  in the range 0.001--0.98. The emergent spectra of all models are redshifted and 
fitted by a diluted blackbody in the  {\it RXTE}/PCA 3--20 keV energy band, and
corresponding  values of the color correction (hardness factors) $\fcol$ are presented. 
}
{Theoretical dependences  $\fcol$--$L/L_{\rm Edd}$ can fitted  to the 
observed dependence $K^{-1/4}$--$F$ of the blackbody normalization $K$ on flux 
during cooling stages of X-ray bursts to determine  
the Eddington flux and the ratio of the apparent neutron star radius to the source distance. 
If the distance is known, these parameters can be transformed to the 
constraints on neutron star mass and radius. 
The theoretical atmosphere spectra  can also be used for direct comparison with the observed 
X-ray burst spectra. }
{}
\keywords{radiative transfer -- scattering --  methods: numerical --
neutron stars -- stars: atmospheres -- X-rays: stars}

\maketitle
%

\section{Introduction}
\label{sec:intro}

Neutron stars (NSs) are the densest  astrophysical objects with apparent surfaces. 
The supra-nuclear density of the NS inner core makes it difficult  to construct the theoretical 
equation of state  (EoS), but it can be constrained if the NS masses and radii are determined 
using astrophysical methods  \citep[see][for a review]{LP07}.  
For example, it is possible in principle to constrain the NS parameters from 
the analysis of the pulse profiles of rapidly rotating NSs -- millisecond  X-ray pulsars  \citep{PG03,LMC08}.  
The apparent NS radius can be obtained from the thermal emission of nearby isolated NSs 
\citep[e.g.][]{Trumperetal:04} or NSs in low-mass X-ray binaries (LMXBs) in globular  clusters during 
quiescence  \citep[e.g.][]{R02,2002ApJ...580..413R,HR06,WB07}. This requires an accurate determination of the distance to the NS
as well as a NS atmosphere model at low effective temperatures to predict the emergent spectrum.

Additional constraints can be obtained for the X-ray bursting NSs.
These are members of LMXBs with a relatively low accretion rate 
showing quasi-periodic X-ray bursts due to  thermonuclear burning of
hydrogen and/or helium at the bottom of the accreted envelope \citep[see e.g.][]{lewin93,SB06}.  
Sometimes these bursts are strong enough and reach the Eddington luminosity $L_{\rm Edd}$. 
In this  case the photosphere radius rapidly increases and the effective temperature decreases 
allowing their easy  identification  \citep{lewin93,GMH08}. 
These photospheric radius expansion (PRE)
bursts provide important information about  the NS
compactness -- the observed Eddington flux and the maximum effective temperature of the NS surface
 \citep[e.g.][]{Ebisuzaki:87,Damen90,vP90}. 
The first one gives  a distance-dependent mass-radius relation, 
while the second one can give a mass--radius relation, which is independent 
of the distance to the NS.   
There exist two uncertainties here. First of all, the chemical composition  of the atmosphere
of X-ray bursting NSs is not well known. Both, the Eddington luminosity and the 
maximum effective temperature depend on it \citep{lewin93}. Second,
X-ray bursting NSs emergent spectra  differ  from the
blackbody at the effective temperature. The emergent
spectra are close to a diluted blackbody with a color temperature  $T_{\rm c}$
larger than the effective temperature by  the color
correction (or hardness) factor $\fcol$. The reason for this is 
Compton scattering in the upper
layers of NS atmospheres during the burst \citep{Londonetal:86, Lapidusetal:86}.  
The value of $\fcol$ depends on the chemical composition and is a strong function of the luminosity 
when it approaches the Eddington limit \citep{Pavlov.etal:91}.
And additionally  more uncertainties arise because the Eddington luminosity depends on the 
gravitational redshift of the photosphere \citep{lewin93}, which varies during the radius expansion phase. 
 
Fast cooling of the NS during the burst with corresponding variations of the effective temperature
allows us to use the whole cooling track to determine NS parameters. 
The burst spectra are usually well described by the blackbody function.   
The normalization of the blackbody $K$  is related to the NS radius $R$ and the distance as
\be \label{u1}
   K \equiv \left( \frac{R_{\rm bb}\,{\rm (km)}}{D_{10}} \right) ^2 = \frac{1}{\fcol ^4} \left( \frac{R\,{\rm (km)}\, (1+z) }{D_{10}}\right) ^2 ,
\ee
where $R_{\rm bb}$ is the blackbody radius, $D_{10}$ is the distance in units of 10 kpc and  $z$ is gravitational
redshift at the NS surface. 
If the photospheric radius coincides with the NS radius, the evolution of  $K$ 
in the cooling tail can only be explained by variations of the color-correction. 
Evolution of $\fcol$   thus should be reflected  
in the evolution of $K^{-1/4}$ with flux \citep{Penninx89,vP90,SP10}. 
Comparing the data with theory, it is possible to determine the Eddington flux and 
the apparent radius $R_{\infty}=R(1+z)$ \citep{SP10}. 
For a NS in a  globular cluster  with known distance  these can be transformed  for the given chemical composition 
into the constraints on  the NS mass and the radius. 

It is clear that this method requires a detailed knowledge of how the 
color-correction varies with luminosity. This behavior depends mainly on the chemical
composition of the  NS atmosphere and less on the surface gravity  $\log g$. 
Spectra of X-ray bursting NSs have been  extensively computed  by different groups since more than two decades
\citep{Londonetal:86, Lapidusetal:86, Ebisuzaki:87, Pavlov.etal:91, madej:91,Madej.etal:04, Madej:05}. 
However, an extensive grid of models is missing. 
Recently, \citet{Madej.etal:04} and \citet{Madej:05} computed 
a set of models, but for a fixed effective temperature and varying $\log g$, while
physically more motivated would be the models with constant $\log g$  and varying relative luminosity $L/L_{\rm Edd}$
 and, therefore, effective temperature.

In the present work we extend previous calculations of NS atmosphere models. 
We consider three realistic NS surface gravities, various chemical compositions, and a broad range of relative luminosities. 
We compute the spectra as well as the color correction factors as a function of  $L/L_{\rm Edd}$. 
We show that the dependence of the blackbody normalization $K^{-1/4}$ on the observed flux can provide information
about the ratio of the NS radius to the distance, the Eddington flux and the chemical composition of  the  atmosphere. 
These can be used to constrain the NS parameters.

\section{Method of calculations}
\label{s:methods}  

We compute  hot X-ray bursting NS model atmospheres in hydrostatic and radiative equilibria 
in plane-parallel approximation. 
Input parameters of the models are the following: chemical
composition (especially important is the hydrogen mass fraction $X$), surface gravity 
\be \label {u2}
    g=\frac{GM}{R^2}(1+z),
\ee
and the relative NS luminosity $l = L/L_{\rm Edd}$.
Here  $R$ and $M$ are the NS radius and the mass, and $L_{\rm Edd}$ is
the Eddington luminosity as measured at the NS surface: 
\be \label {u3}
   L_{\rm Edd} =\frac{4\pi GMc}{\sigma_{\rm e}} (1+z),
\ee
where $\sigma_{\rm e} \approx 0.2(1+X)$ cm$^2$ g$^{-1}$ is the electron
scattering (Thomson) opacity, and 
\be \label {u4}
    1+z=(1-R_{\rm S}/R)^{-1/2} ,
\ee
$R_{\rm S} = 2GM/c^2$ is the Schwarzschild radius of the NS.
The effective temperature $T_{\rm eff}$ can be expressed via $l$ as 
\be \label {u5}
   T_{\rm eff} = l^{1/4} T_{\rm Edd},
\ee
where the Eddington temperature $T_{\rm  Edd}$ is the maximum possible effective temperature on the NS surface:
\be \label {u6}
   \sigma_{\rm SB}T_{\rm Edd}^4 = \frac{gc}{\sigma_{\rm e}}= \frac{GMc}{R^2 \sigma_{\rm e}} (1+z) .
\ee
It is defined from the equality of $g$ and the radiation pressure
acceleration $g_{\rm rad}$ 
\be \label {u7}
g_{\rm rad} =  \frac{4\pi}{c} \int_0^{\infty}  H_{\nu} \, (k_{\nu}+\sigma_{\rm
    e}) \, d\nu \approx \frac{\sigma_{\rm SB} T_{\rm eff}^4 }{c}\sigma_{\rm
    e}
\ee
at the NS surface. Here $k_{\nu}$  is the opacity per unit mass due to free-free and bound-free
transitions (i.e. true opacity, which is much smaller than  electron scattering at
$L \approx L_{\rm Edd}$), and $4\pi H_{\nu}$  is the radiation flux.

The model atmosphere structure for an X-ray bursting NS with given input parameters is described by 
a set of differential equations. The first is the hydrostatic equilibrium
equation  
\be \label{e:hyd}
  \frac {d P_{\rm g}}{dm} = g - g_{\rm rad},
\ee
where  $P_{\rm g}$  is a gas pressure, and the column density $m$ is
determined as
\be
     dm = -\rho \, dz \, ,
\ee
with $\rho$ denoting the gas density and $z$  the vertical distance.

The second is the radiation transfer equation with
Compton scattering  taken into consideration  using the Kompaneets operator 
\citep{Kompaneets:57, Zavlin.Shibanov:91, Grebenev.Sunyaev:02}:
\begin{eqnarray} \label{rtr}
   \frac{\partial^2 f_{\nu} J_{\nu}}{\partial \tau_{\nu}^2} & =& 
\frac{k_{\nu}}{k_{\nu}+\sigma_{\rm e}} \left(J_{\nu} - B_{\nu}\right) -
 \frac{\sigma_{\rm e}}{k_{\nu}+\sigma_{\rm e}} \frac{kT}{m_{\rm e} c^2}
\nonumber  \\
& \times&  x \frac{\partial}{\partial x} \left(x \frac{\partial J_{\nu}}{\partial x} -
3J_{\nu} + \frac{T_{\rm eff}}{T} x J_{\nu} \left[ 1 + \frac{CJ_{\nu}}{x^3}
\right] \right),
\end{eqnarray}
where $x=h \nu /kT_{\rm eff}$ is the dimensionless frequency,
$f_{\nu}(\tau_{\nu}) \approx 1/3$  the variable Eddington factor, $J_{\nu}$ is 
 the mean intensity of radiation, $B_{\nu}$  is the blackbody (Planck)
intensity, $T$  is the local electron temperature,
 and $C=c^2 h^2~/~2(kT_{\rm eff})^3$. The optical
depth $\tau_{\nu}$ is defined as
\be
    d \tau_{\nu} = (k_{\nu}+\sigma_{\rm e}) \, dm.
\ee
These equations are completed  by the energy balance equation
\begin{eqnarray}  \label{econs}
 & & \int_0^{\infty} k_{\nu}\left(J_{\nu} -  B_{\nu}\right) d\nu -
 \sigma_{\rm e} \frac{kT}{m_{\rm e} c^2} \nonumber \\
  & & \times \left( 4 \int_0^{\infty} J_{\nu} \,
d\nu - \frac{T_{\rm eff}}{T} \int_0^{\infty} x J_{\nu}
\left[ 1+\frac{CJ_{\nu}}{x^3}\right] \, d\nu \right)=0,
\end{eqnarray}
the ideal gas law
\be   \label{gstat}
    P_{\rm g} = N_{\rm tot} kT,
\ee
where $N_{\rm tot}$ is the number density of all particles, and also
by the particle and charge conservation equations.  Here we assume local
thermodynamic equilibrium (LTE) in our calculations, so the number
densities of all ionization and excitation states of all elements were
 calculated using Boltzmann and Saha equations, but 
 accounting for the 
 pressure ionization effects  on atomic populations using the occupation
probability formalism \citep{Hum.Mih:88} as  described by \citet{Lanz.Hub:94}.
  We take into account  electron scattering and free-free opacity as well as bound-free transitions  for all ions of the 15 most
abundant chemical elements (H, He, C, N, O, Ne, Na, Mg, Al, Si, S, Ar, Ca, Fe, Ni) \citep[see][]{Ibragimov.etal:03}  using opacities from \cite{VYa:95}.

For solving the above equations, we
used a version of the computer code ATLAS \citep{Kurucz:70,Kurucz:93},
modified to deal with high temperatures.   The code was also
modified to account for Compton scattering \citep{Sul.Pout:06,sw:07}.

The course of calculations is as follows.  First of all, the input
parameters of the model atmosphere  are defined and a
starting model using a grey temperature distribution is calculated.
The calculations are performed with a set of 98 depth points $m_{\rm
i}$ distributed logarithmically in equidistant steps from $m\approx
10^{-7}$~g~cm$^{-2}$ to $m_{\rm max}=10^6$~g~cm$^{-2}$. The appropriate value of
$m_{\rm max}$ is such that satisfies the condition $\sqrt{\tau_{\nu,\rm
b-f,f-f}(m_{\rm max})\tau_{\nu}(m_{\rm max})} >$ 1 at all frequencies.
Here $\tau_{\nu,\rm b-f,f-f}$ is the optical depth computed with the true
opacity  only (bound-free and free-free transitions, without scattering).  
Satisfying this equation is necessary for the inner boundary condition
of the radiation transfer problem.

We used the condition of the absence of irradiation flux at the outer boundary
\be
    \frac{\partial J_{\nu}}{\partial \tau_{\nu}} = h_{\nu} J_{\nu},
\ee
where $h_{\nu}$ is the surface variable Eddington factor.
The inner boundary condition, 
\be
   \frac{\partial J_{\nu}}{\partial \tau_{\nu}} =
 \frac{\partial B_{\nu}}{\partial \tau_{\nu}},
\ee
 is obtained
from the diffusion approximation $J_{\nu} \approx B_{\nu}$ and $H_{\nu}
\approx 1/3 \times \partial B_{\nu}/\partial \tau_{\nu}$.

The boundary conditions along the frequency axis are
\be  \label{lbc}
      J_{\nu} = B_{\nu}
\ee
at the lower frequency boundary ($\nu_{\min}=10^{14}$ Hz,
 $h\nu_{\min} \ll kT_{\rm eff}$), and
\be  \label{hbc}
x \frac{\partial J_{\nu}}{\partial x} - 3J_{\nu} + \frac{T_{\rm eff}}{T} x
J_{\nu} \left( 1 + \frac{CJ_{\nu}}{x^3} \right)=0
\ee
at the upper frequency boundary ($\nu_{\max}\approx  
10^{19}$ Hz, $h\nu_{\max} \gg kT_{\rm eff}$).  Condition
(\ref{lbc}) means that at the lowest energies the true opacity
dominates over scattering $k_{\nu} \gg \sigma_{\rm e}$, and therefore
$J_{\nu} \approx B_{\nu}$. Condition (\ref{hbc}) means that there is
no photon flux along the frequency axis at the highest energy.

 For the starting model, all number densities and opacities at all
depth points and all frequencies are calculated. We use 300 logarithmically
equidistant frequency points in our computations. The radiation transfer
equation (\ref{rtr}) is non-linear and is solved iteratively by the
Feautrier method \citep[][~see also \citealt{Zavlin.Shibanov:91,
Pavlov.etal:91,Grebenev.Sunyaev:02}]{Mihalas:78}.  We use the last
term of Eq. (\ref{rtr}) in the form
$xJ_{\nu}^i(1+CJ_{\nu}^{i-1}/x^3)$, where $J_{\nu}^{i-1}$ is the mean
intensity from the previous iteration.  During the first iteration we
take $J_{\nu}^{i-1}=0$.  Between iterations we calculate the variable
Eddington factors $f_{\nu}$ and $h_{\nu}$, using the formal solution
of the radiation transfer equation in three angles at each frequency.
We use six iterations, because usually in the considered models  4--5 iterations
are sufficient for  achieving convergence.

The solution of the radiative transfer equation (\ref{rtr}) is checked
for the energy balance equation (\ref{econs}), together with the surface
flux condition
\be
    4 \pi \int_0^{\infty} H_{\nu} (m=0) d\nu = \sigma_{\rm SB} T_{\rm eff}^4 = 4 \pi
   H_0.
\ee
The relative flux error 
\be
     \varepsilon_{H}(m) = 1 - \frac{H_0}{\int_0^{\infty} H_{\nu} (m) d\nu},
\ee
and the energy balance error as  functions of depth
\begin{eqnarray}  \label{econs1}
 \varepsilon_{\Lambda}(m)&=&  \int_0^{\infty} k_{\nu}\left(J_{\nu} - B_{\nu}\right) d\nu -
 \sigma_{\rm e} \frac{kT}{m_{\rm e} c^2} \nonumber \\
& \times& \left( 4 \int_0^{\infty} J_{\nu} \,
d\nu - \frac{T_{\rm eff}}{T} \int_0^{\infty} x J_{\nu}
\left[1+\frac{CJ_{\nu}}{x^3}\right] \, d\nu \right)
\end{eqnarray}
are calculated, where $H_{\nu} (m)$ is the radiation flux at any given depth $m$.
This quantity is found from
the first moment of the radiation transfer equation:
\be
    \frac{\partial f_{\nu} J_{\nu}}{\partial \tau_{\nu}} = H_{\nu}.
\ee
Temperature corrections are then evaluated using three different
procedures.  The first is the integral $\Lambda$-iteration method,
modified for Compton scattering, based on the energy balance equation
(\ref{econs}). In this method the temperature correction for a particular
depth is found from 
\be
     \Delta T_{\Lambda} = \frac{-\varepsilon_{\Lambda}(m)}{\int_0^{\infty}
\left[ (\Lambda_{\nu\,{\rm diag}}-1)/(1-\alpha_{\nu}\Lambda_{\nu\,{\rm diag}}) \right]
k_{\nu} (dB_{\nu}/dT)\,d \nu}.
\ee   
Here $\alpha_{\nu}=\sigma_{\rm e}/(k_{\nu}+\sigma_{\rm e})$, and
$\Lambda_{\nu\,{\rm diag}}$  is the diagonal matrix element of the $\Lambda$
operator.
This procedure is used in the upper atmospheric layers.  The
second procedure is the Avrett-Krook flux correction, which uses the
relative flux error $\varepsilon_{H}(m)$ and is performed in the deep layers.
And the third 
one is the surface correction, which is based on the emergent flux
error.  See \citet{Kurucz:70} for a detailed description of the 
methods.

The iteration procedure is repeated until the relative flux error is
smaller than 1\%, and the relative flux derivative error is smaller
than 0.01\%. As a result of these calculations, we obtain a
self-consistent X-ray bursting NS  model atmosphere, together with the emergent
spectrum of radiation.

This code was tested against the code by \citet{Madej.etal:04} on 
problems of the atmospheres of X-ray bursting NSs and hot DA white dwarfs by 
\citet{Sul.Pout:06}  and \citet{Sul2006}, respectively. 
See detailed discussion in Sect. \ref{sec:compar}.

\section{Results of atmosphere modeling}
\label{sec:model}

\begin{figure}
\begin{center}
\includegraphics[width=0.92\columnwidth]{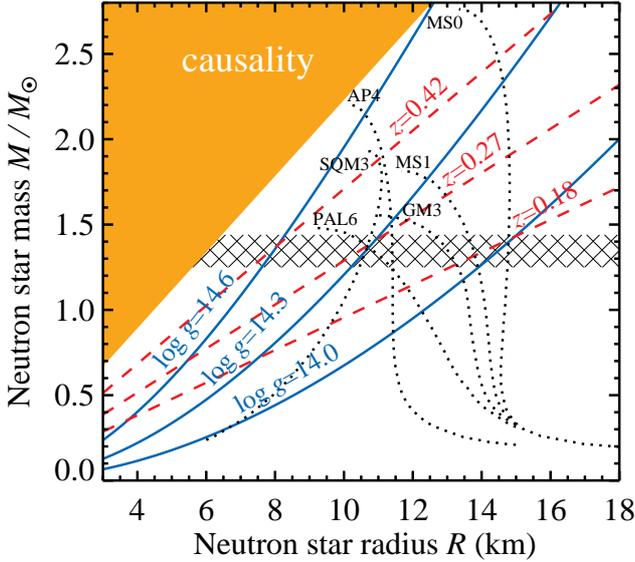}
\caption{\label{fig:logg}
The mass-radius relations for several equations of state of neutron and strange stars matter are shown by dotted curves. 
The $M$--$R$ relations for constant $\log g=14.0, 14.3, 14.6$ are shown by solid curves and 
the lines of constant redshift ($z=0.18, 0.27, 0.42$ corresponding to the chosen $\log g$ for $M=1.4M_{\odot}$) by dashed curves. 
The upper-left region is excluded  by constraints from the causality requirements \citep{HPY07,LP07}.
The hatched horizontal belt marks the spread of pulsars masses accurately measured in double NS binaries \citep{HPY07}. 
}
\end{center}
\end{figure}

\begin{figure}
\begin{center}
\includegraphics[width=1.0\columnwidth]{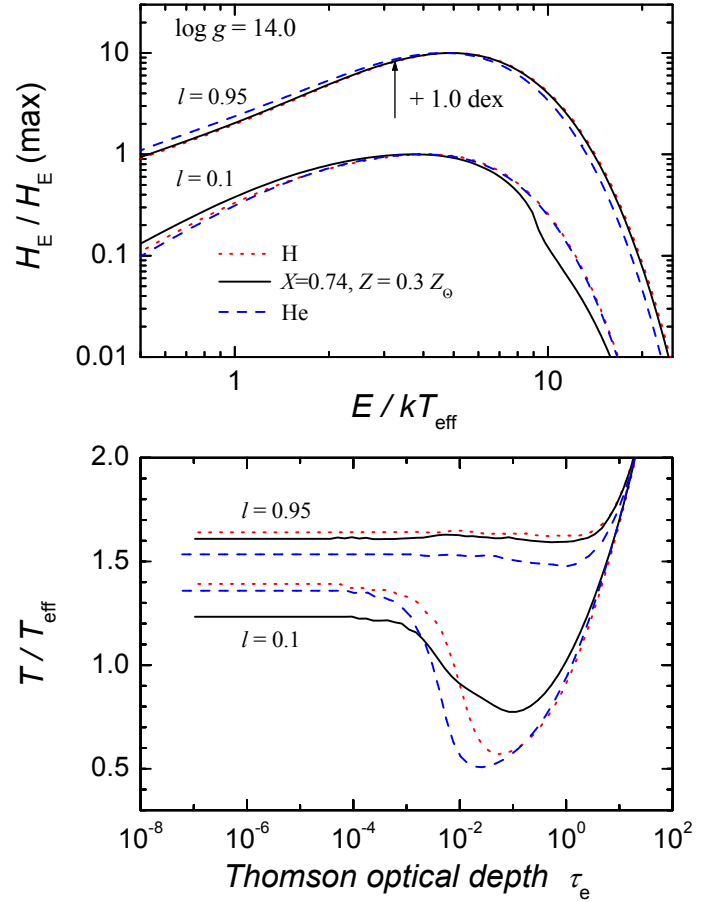}
\caption{\label{fig:hhes}
Emergent (unredshifted) spectra and temperature structures of  model atmospheres with
two relative luminosities ($l$ = 0.95 and 0.1) at fixed surface gravity ($\log
g$ = 14.0) for various chemical compositions: pure hydrogen (dotted
curves), pure helium (dashed curves) and  
solar mix of  hydrogen and  helium, and subsolar metal abundance $Z=0.3Z_{\odot}$ (solid curves).  
{\it Top panel:}  Spectra normalized to maximum flux versus photon energy in units of the 
effective temperature. Note, that $T_{\rm eff}\propto l^{1/4} (1+X)^{-1/4}$.
For clarity, the spectra for $l=0.95$ are shifted along the ordinate axis by a factor of 10.
{\it Bottom panel:} Corresponding temperature structures (in units  of the effective
temperature) versus Thomson optical depth.
}
\end{center}
\end{figure}

\begin{figure}
\begin{center}
\includegraphics[width=1.0\columnwidth]{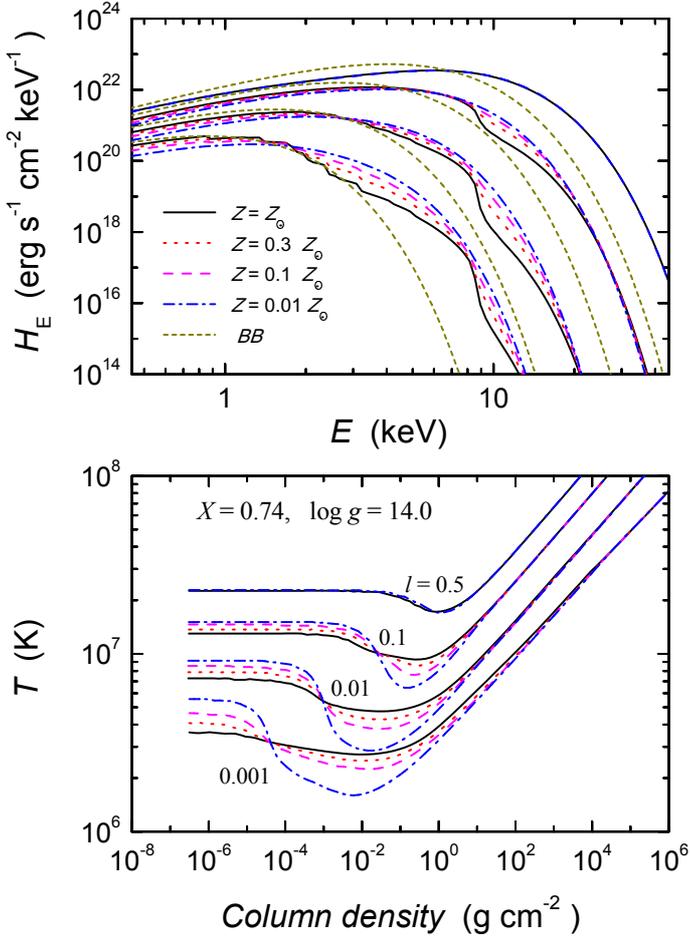}
\caption{\label{fig:fe}
Emergent (unredshifted) spectra ({\it top panel}) and temperature structures ({\it bottom
  panel}) of the model atmospheres with
four relative luminosities ($l$ = 0.5, 0.1, 0.01 and 0.001) and fixed surface gravity ($\log
g$ = 14.0) for solar hydrogen-helium mixture and various
abundances of heavy elements: $Z=Z_{\odot}$ (solid curves), $Z=0.3Z_{\odot}$ (dotted
curves), $Z=0.1Z_{\odot}$  (dashed curves), $Z=0.01Z_{\odot}$  (dot-dashed curves). 
In the top panel, the blackbody spectra with effective temperatures are also shown by short-dashed curves.
}
\end{center}
\end{figure}

\begin{figure}
\begin{center}
\includegraphics[width=1.0\columnwidth]{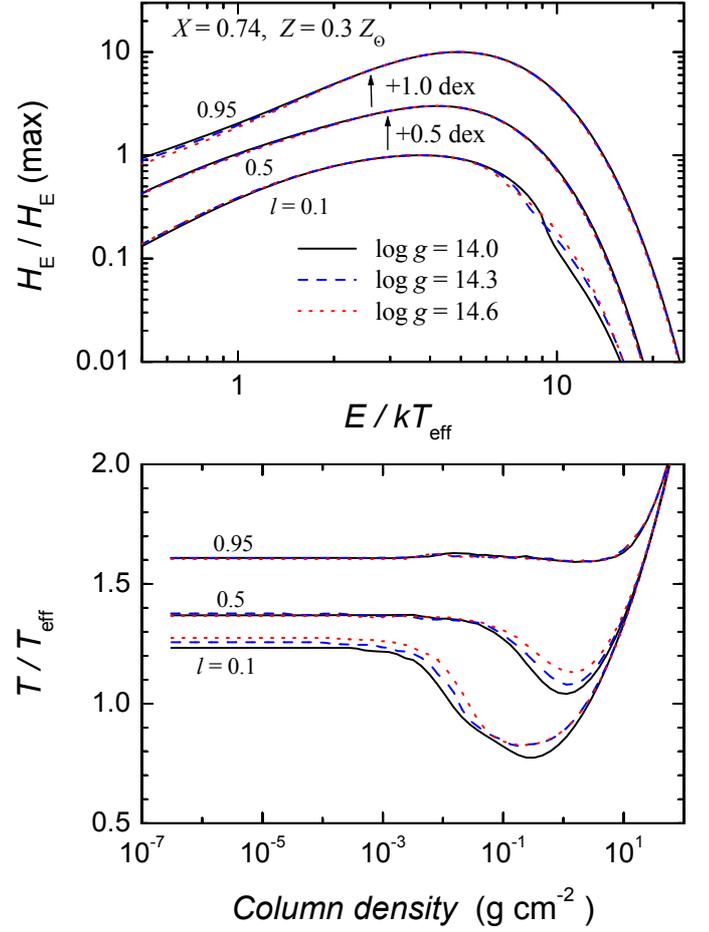}
\caption{\label{fig:g}
Emergent (unredshifted) spectra ({\it top panel}) and temperature structures ({\it bottom
  panel}) for three relative luminosities ($l$ = 0.95, 0.5 and 0.1)
and fixed chemical abundance (solar hydrogen-helium mixture and  $Z=0.3Z_{\odot}$) 
for three different surface gravities: $\log g$ = 14.0 (solid curves), 14.3
(dashed curves) and 14.6 (dotted curves). For clarity, the spectra for
$l$ =0.95 and 0.5 are
shifted along the ordinate axis by factors 10$^{+1.0}$ and 10$^{+0.5}$, respectively.
}
\end{center}
\end{figure}

Using the code described above we have calculated an extended set of the hot
NS atmospheres. The model atmospheres were computed for three values
of surface gravity $\log g$ = 14.0, 14.3 and 14.6, which cover most of the physically realistic NS equations of state
for a large range of NS masses (see Fig. \ref{fig:logg}). 
We consider six chemical compositions: pure hydrogen, pure helium, and a solar mixture of
the hydrogen and helium with various abundances of  heavy elements scaled
to solar abundances, $Z$ = 1, 0.3, 0.1 and 0.01 times the solar ($Z_\odot$), or using the standard
stellar atmosphere definition [Fe/H] = 0, $-$0.5, $-$1 and $-$2. We used new values
of the solar abundances \citep{Asplundetal:09}. In particular, according to
this work the number ratio of helium to hydrogen is less than was
adopted before ($n$(He)/$n$(H) = 0.0851 instead of 0.0977, a new hydrogen mass
fraction $X$ = 0.7374).
For every values of
abundance and $\log g$ twenty models with relative luminosity $l$ = 0.001, 0.003, 
0.01, 0.03, 0.05, 0.07,  0.1, 0.15, 0.2, 0.3, 0.4, 0.5, 0.6, 0.7, 0.75, 0.8, 0.85, 0.9, 0.95, 0.98 
were calculated. The total number of the
computed models in the set is 360.
Basic properties of the obtained models are shown in Figs.\,\ref{fig:hhes}--\ref{fig:g}. 
Most of them are well known from the previous investigations 
(see Section~\ref{sec:intro}) and we present these figures mainly for illustrative purposes.

The effect of chemical composition on model emergent spectra and
temperature structures is illustrated in Fig.\,\ref{fig:hhes}.
Normalized spectra and temperature distributions for models with fixed $\log
g$ = 14.0 and relatively high ($l$ = 0.95) and low ($l$ = 0.1) luminosities
for three different chemical compositions (pure hydrogen, pure helium and
solar abundances) are shown. The dependences of the normalized spectra on
the photon energy scaled to the effective temperatures are presented.

Atmosphere temperature structures in optically
thin layers (at electron scattering optical depth $\tau_{\rm e} < 1$) 
are determined by the balance between 
heating of electrons due to down-scattering of high energy photons from deep layers and 
 cooling due to free-free and bound-free emission. 
 In the upper low-density layers, cooling is inefficient and therefore 
 the temperature rises and forms a chromosphere-like structure in less luminous models. 
In the high luminosity atmospheres,  radiation pressure is significant and the plasma density is
low. Due to these conditions, the extended high temperature layers stretch up
to optically thick layers because of the influence of Compton heating.

Emergent spectra of high temperature NS atmospheres are close to diluted blackbody
spectra \citep{Londonetal:86, Lapidusetal:86}  
\be
\label{u14}
   F_{\rm E} = \frac{1}{\fcol^4} B_{\rm E} (T_{\rm c} = \fcol T_{\rm eff})
\ee
with color correction (or hardness) factor $\fcol$ (see Section \ref{sec:results} for
details). To first approximation $\fcol$ is equal to the ratio of the upper layers
(surface) 
temperature to $T_{\rm eff}$. Cooling is more effective in pure helium
atmospheres, therefore they have lower surface temperatures and relatively
soft spectra (smaller $\fcol$). 

In the low luminosity atmospheres with heavy
elements, iron is not completely ionized. Therefore, an absorption edge at 9
keV due to the bound-free transition from the ground level of the H-like Fe {\sc xxvi}  ion
arises. This edge reduces the number of hard photons, therefore, the upper
layers stay at lower temperatures. The strength of this edge and the surface
temperature depend on the metal abundance as demonstrated in Fig.\,\ref{fig:fe}. At the lowest
luminosity,  absorption edges in the 1--3 keV range due to other chemical elements arise. It is
interesting that emergent spectra and temperature structures of the more luminous
models ($l \ge 0.5$) depend on the heavy element abundances very little,
because at these high temperatures iron is completely ionized.

Emergent spectra and temperature structures for three relative luminosities 
and for three different surface gravities  are shown in Fig.\,\ref{fig:g} for 
fixed chemical abundances (solar H/He mix and $Z=0.3Z_{\odot}$). 
 Again, the luminous models for a given $l$ depend on $\log g$ very
little, but for the low luminosity models this dependence is significant. The
model with lowest gravity has the lowest $T_{\rm eff}$ and the lowest iron
ionization. It leads to the largest absorption edge and the lowest surface temperature.

\begin{figure}
\begin{center}
\includegraphics[width=0.9\columnwidth]{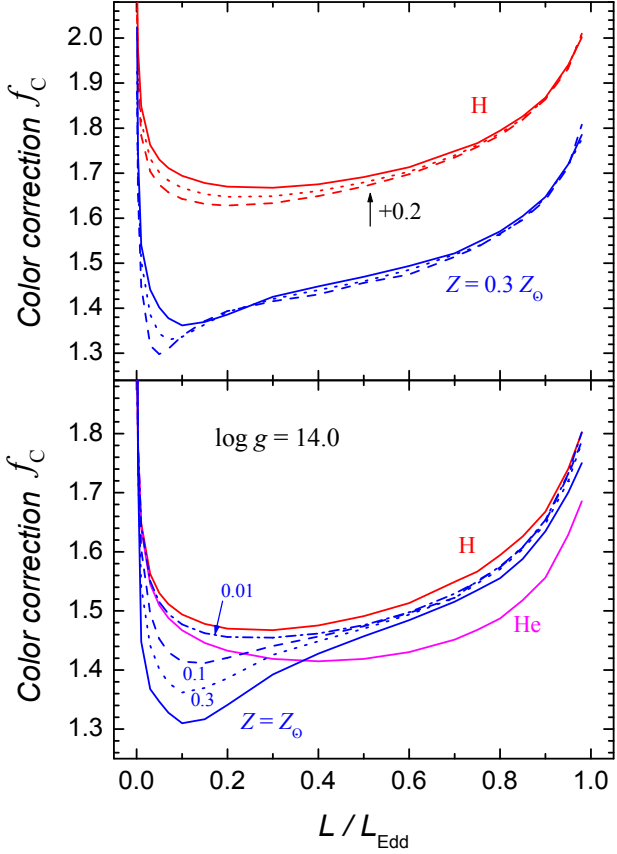}
\caption{\label{fig:fc_main}
Dependence of the color correction factors $\fcol$ on the relative luminosity 
for various NS atmosphere models. 
The $\fcol$ are obtained using the first fitting procedure.
{\it Top panel:} Dependences for  hydrogen
and solar H/He mixture with $Z=0.3Z_{\odot}$ models with different surface
gravities $\log g$ = 14.0 (solid curves), 14.3 (dotted curves) and 14.6
(dashed curves). For clarity, the dependences for hydrogen models are shifted up by +0.2.
{\it Bottom panel:} Variation of $\fcol$ on chemical compositions: pure hydrogen (upper curve), pure helium
(lowest curve), and solar H/He mixture with $Z=Z_{\odot}$ (solid curves), $Z=0.3Z_{\odot}$ (dotted
curves), $Z=0.1Z_{\odot}$  (dashed curves), $Z=0.01Z_{\odot}$  (dot-dashed curves) for low gravity $\log g$ = 14.0 models.  
 }
\end{center}
\end{figure}

\section{Color correction factor}
\label{sec:results}

For a given model atmosphere  defined by $l$, $\log g$ and  chemical composition, we compute also 
the emergent spectra. These spectra are then fitted by 
a diluted blackbody spectrum (\ref{u14}) 
in some energy band, for example, 3--20 keV, corresponding to the very often
used PCA detector of the {\it RXTE} observatory. 
As in the case of the observations, we use the fits with two free parameters, the normalization $w$ 
and the color-correction:   
\be
\label{u15}
   F_{E} \approx w B_{E} (\fcol T_{\rm eff}).
\ee
This two-parameter approximation gives rather good fits, especially for low-temperature  
atmospheres with heavy elements (see Fig.\,\ref{fig:fits}).
The difference between $w$ and $\fcol^{-4}$ is small for luminous models.

Results of the fitting depend on the used fitting procedure, as was
first pointed out by \cite{Ebisuzaki:87}. 
Here we use five different procedures. 
In the first one, we minimize the sum
\be
\label{u16}
   \sum^N_{n=1} (F_{E_n} - w_1 B_{E_n} (f_{\rm c,1} T_{\rm eff}))^2,
\ee
where $N$ is the number of photon energy points in the considered energy band. 
As the energy points in the computed spectra are equidistant in logarithm, 
this procedure is formally equivalent to minimizing the integral
\be
\label{u17}
   \int^{E_{\rm max}}_{E_{\rm min}} (F_{E} - w_1 B_{E} (f_{\rm c,1} T_{\rm eff}))^2\,\frac{dE}{E}.
\ee
When fitting the data, one fits the photon count flux, not the energy flux, therefore, 
in the second procedure we minimize the following sum
\be
\label{u18}
   \sum^N_{n=1} \frac{(F_{E_n} - w_2 B_{E_n} (f_{\rm c,2} T_{\rm eff}))^2}{E^2_n},
\ee
which is  equivalent to minimizing the integral
\be
\label{u19}
   \int^{E_{\rm max}}_{E_{\rm min}} (F_{E} - w_2 B_{E} (f_{\rm c,2} T_{\rm eff}))^2\,\frac{dE}{E^3}. 
\ee
In the third procedure we suggest to minimize the integral
\be
\label{u20}
   \int^{E_{\rm max}}_{E_{\rm min}} (F_{E} - w_3 B_{E} (f_{\rm c,3} T_{\rm eff}))^2\,dE,
\ee
which corresponds to minimizing the sum
\be
\label{u21}
   \sum^N_{n=1} (F_{E_n} - w_3 B_{E_n} (f_{\rm c,3} T_{\rm eff}))^2  E_n .
\ee 
We also use a fourth fit with only one free parameter, relating $w = f_{\rm c,4}^{-4}$, and using the same minimization as for the first 
procedure. 
And finally in the fifth procedure we compute the color correction $f_{\rm c,5}$ by dividing the energy where the peak of the model flux $F_{E}$
is reached by the peak energy of the blackbody spectrum $B_{E} (T_{\rm eff})$ as was done by \citet{Madej.etal:04} and \citet{Madej:05}.

\begin{figure}
\begin{center}
\includegraphics[width=0.9\columnwidth]{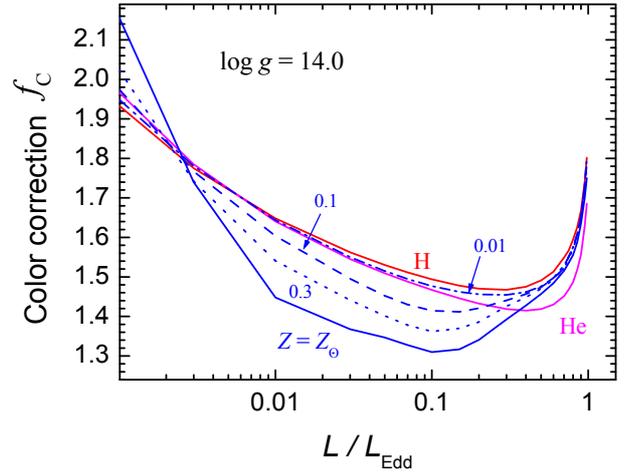}
\caption{\label{fig:fc_log}
Same as bottom panel of Fig. \ref{fig:fc_main}, but in a log-scale.  }
\end{center}
\end{figure}

\begin{figure}
\begin{center}
\includegraphics[width=0.9\columnwidth]{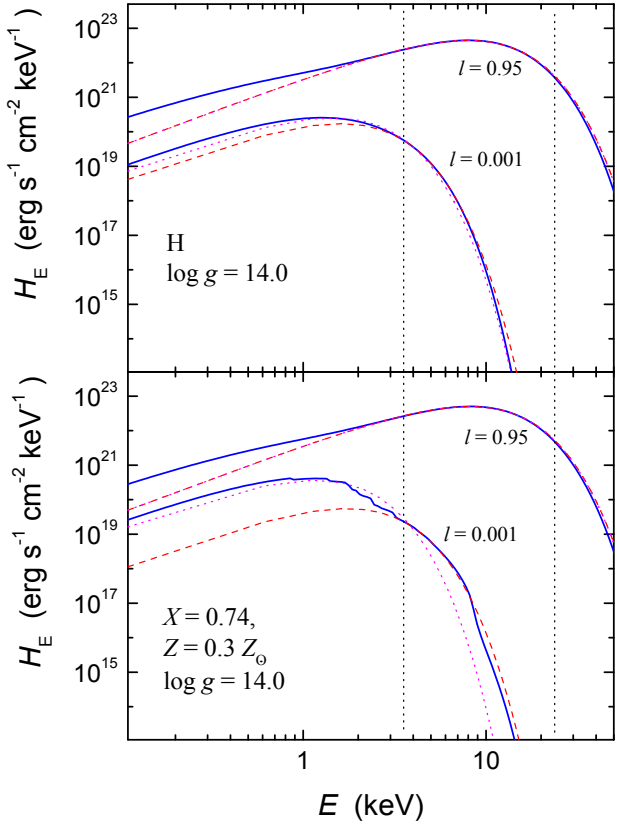}
\caption{\label{fig:fits}
Examples  NS atmosphere spectra 
for  high ($l$=0.95) and low ($l$=0.001) luminosity and low gravity ($\log g$ = 14.0). 
Theoretical model spectra are shown by  the solid curves.
The diluted blackbody  fits with the first fit procedure are shown by the dashed curves
and the one-parameter fits  (fourth fit procedure) are shown by the dotted curves. 
The vertical dotted lines show boundaries of the energy band where the fitting
procedure was performed.
{\it Top panel} is for  hydrogen atmosphere and the {\it bottom panel} is for the solar H/He mixture with $Z=0.3Z_{\odot}$.
 }
\end{center}
\end{figure}

\begin{figure}  
\centering
\includegraphics[width=0.83\columnwidth]{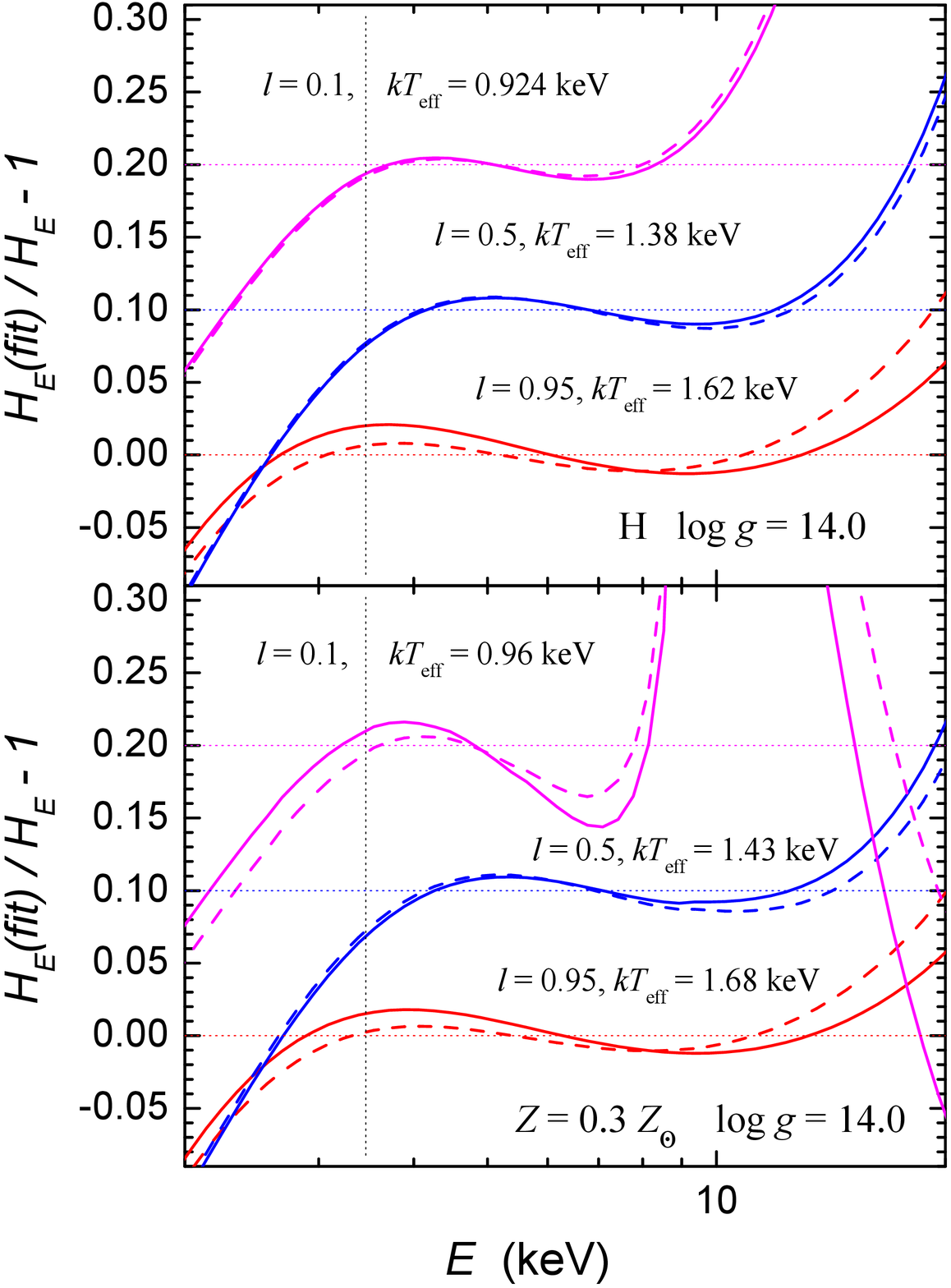}
\caption{\label{fig:resid}
Relative errors of the computed spectra fitted by the first (solid curves) and the second
(dashed curves) fit procedures vs. photon energy for hydrogen ({\it top panel})
and solar H/He mixture with $Z=0.3Z_{\odot}$ ({\it bottom panel}) low gravity
($\log g$ =14.0) models. Corresponding relative luminosities and effective
temperatures are given at the curves. The vertical dotted line shows the low boundary
of the energy band where  fitting procedures were performed. For clarity,
models with $l$ = 0.5 and 0.1 were shifted by +0.1 and +0.2, respectively.  
 }
\end{figure}

The obtained color correction factors $f_{\rm c,1}$--$f_{\rm c,4}$ depend on the chosen energy band. 
We  perform our four fitting procedures and calculate the corresponding color correctios and dilution factors
in the 3--20 keV energy band corresponding to the {\it RXTE}/PCA detector. 
Due to gravitational redshift a spectrum in the observed energy
band is radiated in the energy band with blueshifted boundaries $(3 - 20)\times(1+z)$ keV. 
Each NS has its own (a priori unknown) gravitational redshift. 
We calculated redshifts using $\log g$ and adopting a NS mass equal to
1.4 $M_{\odot}$ (see Eqs.\,(\ref{u2}) and (\ref{u4})). We obtained $R$ = 14.80,
10.88, 8.16 km  and   $z$ = 0.18, 0.27, 0.42 for $\log g$ = 14.0, 14.3, and 14.6, respectively.
Varying the mass in the interval $M_{\odot}$--2$M_{\odot}$ introduces 3, 5, 10\%  uncertainties in $1+z$ 
(see Fig. \ref{fig:logg}), which, however, have less  than a 0.1\% effect on the color corrections.
The results of the fitting procedures are presented in Table\, \ref{tab:fc} and in the online material.

\begin{table*}
\caption{\label{tab:fc}
Color-correction and dilution factors from the blackbody fits to the spectra of hydrogen 
atmosphere  models at $\log g=14.0$. 
The results for other chemical compositions and gravities  are presented in the Online Material.} 
\begin{center}
\begin{tabular}{lccccccccc}
\hline 
\hline \multicolumn{10}{c}{Set  1\,\,\,\,                $X=1$ $\quad$ $Z=0$  $\,\,\,\, \log g = 14.0 \,\,\,\, T_{\rm Edd}= 1.644$ keV  $\quad R=   14.80$ km  $\quad  z=   0.18$} \\
\hline
   $l$       &   $T_{\rm eff}$ (keV)   &   $f_{\rm c,1}$ &  $f_{\rm c,2}$    & $f_{\rm c,3}$ &       
    $f_{\rm c,4}$ & $f_{\rm c,5}$ & $w_1 f_{\rm c,1}^4$  & $w_2 f_{\rm c,2}^4$ &  $w_3 f_{\rm c,3} ^4 $ \\
\hline
  0.001 &   0.292 &   1.933 &   1.932 &   1.934 &   1.765 &   1.538 &   0.742 &   0.743 &   0.741  \\ 
  0.003 &   0.385 &   1.775 &   1.774 &   1.775 &   1.638 &   1.448 &   0.829 &   0.830 &   0.829  \\ 
  0.010 &   0.520 &   1.648 &   1.649 &   1.647 &   1.543 &   1.382 &   0.900 &   0.900 &   0.900  \\ 
  0.030 &   0.684 &   1.563 &   1.565 &   1.560 &   1.520 &   1.396 &   0.942 &   0.941 &   0.942  \\ 
  0.050 &   0.777 &   1.530 &   1.533 &   1.528 &   1.523 &   1.429 &   0.954 &   0.954 &   0.955  \\ 
  0.070 &   0.845 &   1.511 &   1.514 &   1.509 &   1.517 &   1.448 &   0.961 &   0.961 &   0.961  \\ 
  0.100 &   0.924 &   1.494 &   1.497 &   1.491 &   1.506 &   1.461 &   0.966 &   0.967 &   0.966  \\ 
  0.150 &   1.023 &   1.478 &   1.480 &   1.475 &   1.494 &   1.469 &   0.970 &   0.971 &   0.970  \\ 
  0.200 &   1.099 &   1.470 &   1.472 &   1.467 &   1.487 &   1.471 &   0.972 &   0.973 &   0.972  \\ 
  0.300 &   1.217 &   1.467 &   1.467 &   1.465 &   1.484 &   1.479 &   0.975 &   0.975 &   0.975  \\ 
  0.400 &   1.307 &   1.475 &   1.474 &   1.473 &   1.491 &   1.491 &   0.977 &   0.976 &   0.976  \\ 
  0.500 &   1.382 &   1.491 &   1.489 &   1.489 &   1.506 &   1.511 &   0.979 &   0.978 &   0.979  \\ 
  0.600 &   1.447 &   1.513 &   1.511 &   1.511 &   1.528 &   1.535 &   0.980 &   0.978 &   0.979  \\ 
  0.700 &   1.504 &   1.549 &   1.547 &   1.547 &   1.560 &   1.571 &   0.986 &   0.984 &   0.985  \\ 
  0.750 &   1.530 &   1.567 &   1.566 &   1.564 &   1.578 &   1.589 &   0.984 &   0.984 &   0.983  \\ 
  0.800 &   1.555 &   1.594 &   1.596 &   1.591 &   1.601 &   1.617 &   0.992 &   0.993 &   0.991  \\ 
  0.850 &   1.578 &   1.626 &   1.630 &   1.622 &   1.631 &   1.647 &   0.994 &   0.997 &   0.993  \\ 
  0.900 &   1.601 &   1.668 &   1.678 &   1.662 &   1.672 &   1.688 &   0.994 &   1.000 &   0.993  \\ 
  0.950 &   1.623 &   1.740 &   1.759 &   1.731 &   1.737 &   1.758 &   1.004 &   1.015 &   1.001  \\ 
  0.980 &   1.635 &   1.802 &   1.832 &   1.788 &   1.809 &   1.815 &   0.992 &   1.009 &   0.989  \\  
   \hline
\end{tabular}
\end{center}
\end{table*}

\onllongtab{1}{
\label{tab:online}
\begin{longtable}{cccccccccc} 
\caption{Color-correction and dilution factors from the blackbody fits to the atmosphere model spectra. } \\
\hline 
\hline
   $l$       &   $T_{\rm eff}$ (keV)   &   $f_{\rm c,1}$ &  $f_{\rm c,2}$    & $f_{\rm c,3}$ &       
    $f_{\rm c,4}$ & $f_{\rm c,5}$ & $w_1 f_{\rm c,1}^4$  & $w_2 f_{\rm c,2}^4$ &  $w_3 f_{\rm c,3} ^4 $ \\
\hline
\endfirsthead
\caption{Continued.} \\
\hline
  $l$       &   $T_{\rm eff}$ (keV)   &   $f_{\rm c,1}$ &  $f_{\rm c,2}$    & $f_{\rm c,3}$ &       
   $f_{\rm c,4}$ & $f_{\rm c,5}$ & $w_1 f_{\rm c,1}^4$  & $w_2 f_{\rm c,2}^4$ &  $w_3 f_{\rm c,3} ^4 $ \\
  \hline
\endhead
\hline
\endfoot
\hline
\endlastfoot
 \hline \multicolumn{10}{c}{Set  1\,\,\,\,                $X=1$ $\quad$ $Z=0$  $\,\,\,\, \log g = 14.0 \,\,\,\, T_{\rm Edd}= 1.644$ keV  $\quad R=   14.80$ km  $\quad  z=   0.18$} \\ \hline
  0.001 &   0.292 &   1.933 &   1.932 &   1.934 &   1.765 &   1.538 &   0.742 &   0.743 &   0.741  \\ 
  0.003 &   0.385 &   1.775 &   1.774 &   1.775 &   1.638 &   1.448 &   0.829 &   0.830 &   0.829  \\ 
  0.010 &   0.520 &   1.648 &   1.649 &   1.647 &   1.543 &   1.382 &   0.900 &   0.900 &   0.900  \\ 
  0.030 &   0.684 &   1.563 &   1.565 &   1.560 &   1.520 &   1.396 &   0.942 &   0.941 &   0.942  \\ 
  0.050 &   0.777 &   1.530 &   1.533 &   1.528 &   1.523 &   1.429 &   0.954 &   0.954 &   0.955  \\ 
  0.070 &   0.845 &   1.511 &   1.514 &   1.509 &   1.517 &   1.448 &   0.961 &   0.961 &   0.961  \\ 
  0.100 &   0.924 &   1.494 &   1.497 &   1.491 &   1.506 &   1.461 &   0.966 &   0.967 &   0.966  \\ 
  0.150 &   1.023 &   1.478 &   1.480 &   1.475 &   1.494 &   1.469 &   0.970 &   0.971 &   0.970  \\ 
  0.200 &   1.099 &   1.470 &   1.472 &   1.467 &   1.487 &   1.471 &   0.972 &   0.973 &   0.972  \\ 
  0.300 &   1.217 &   1.467 &   1.467 &   1.465 &   1.484 &   1.479 &   0.975 &   0.975 &   0.975  \\ 
  0.400 &   1.307 &   1.475 &   1.474 &   1.473 &   1.491 &   1.491 &   0.977 &   0.976 &   0.976  \\ 
  0.500 &   1.382 &   1.491 &   1.489 &   1.489 &   1.506 &   1.511 &   0.979 &   0.978 &   0.979  \\ 
  0.600 &   1.447 &   1.513 &   1.511 &   1.511 &   1.528 &   1.535 &   0.980 &   0.978 &   0.979  \\ 
  0.700 &   1.504 &   1.549 &   1.547 &   1.547 &   1.560 &   1.571 &   0.986 &   0.984 &   0.985  \\ 
  0.750 &   1.530 &   1.567 &   1.566 &   1.564 &   1.578 &   1.589 &   0.984 &   0.984 &   0.983  \\ 
  0.800 &   1.555 &   1.594 &   1.596 &   1.591 &   1.601 &   1.617 &   0.992 &   0.993 &   0.991  \\ 
  0.850 &   1.578 &   1.626 &   1.630 &   1.622 &   1.631 &   1.647 &   0.994 &   0.997 &   0.993  \\ 
  0.900 &   1.601 &   1.668 &   1.678 &   1.662 &   1.672 &   1.688 &   0.994 &   1.000 &   0.993  \\ 
  0.950 &   1.623 &   1.740 &   1.759 &   1.731 &   1.737 &   1.758 &   1.004 &   1.015 &   1.001  \\ 
  0.980 &   1.635 &   1.802 &   1.832 &   1.788 &   1.809 &   1.815 &   0.992 &   1.009 &   0.989  \\ 
\hline \multicolumn{10}{c}{Set  2\,\,\,\,                $X=1$ $\quad$ $Z=0$  $\,\,\,\, \log g = 14.3 \,\,\,\, T_{\rm Edd}= 1.954$ keV  $\quad R=   10.88$ km  $\quad  z=   0.27$} \\ \hline
  0.001 &   0.347 &   1.889 &   1.888 &   1.890 &   1.735 &   1.531 &   0.777 &   0.779 &   0.776  \\ 
  0.003 &   0.457 &   1.737 &   1.736 &   1.737 &   1.611 &   1.443 &   0.855 &   0.857 &   0.855  \\ 
  0.010 &   0.618 &   1.615 &   1.615 &   1.614 &   1.525 &   1.382 &   0.916 &   0.916 &   0.917  \\ 
  0.030 &   0.813 &   1.532 &   1.534 &   1.530 &   1.513 &   1.403 &   0.951 &   0.951 &   0.952  \\ 
  0.050 &   0.924 &   1.501 &   1.504 &   1.499 &   1.505 &   1.430 &   0.961 &   0.961 &   0.962  \\ 
  0.070 &   1.005 &   1.484 &   1.486 &   1.481 &   1.496 &   1.442 &   0.966 &   0.967 &   0.966  \\ 
  0.100 &   1.099 &   1.467 &   1.470 &   1.464 &   1.482 &   1.449 &   0.970 &   0.971 &   0.970  \\ 
  0.150 &   1.216 &   1.453 &   1.455 &   1.451 &   1.469 &   1.452 &   0.973 &   0.974 &   0.973  \\ 
  0.200 &   1.306 &   1.448 &   1.448 &   1.445 &   1.464 &   1.454 &   0.975 &   0.975 &   0.974  \\ 
  0.300 &   1.446 &   1.449 &   1.447 &   1.447 &   1.464 &   1.463 &   0.977 &   0.975 &   0.976  \\ 
  0.400 &   1.554 &   1.460 &   1.456 &   1.459 &   1.475 &   1.480 &   0.979 &   0.976 &   0.978  \\ 
  0.500 &   1.643 &   1.479 &   1.475 &   1.478 &   1.493 &   1.500 &   0.981 &   0.978 &   0.981  \\ 
  0.600 &   1.719 &   1.503 &   1.499 &   1.502 &   1.518 &   1.526 &   0.980 &   0.978 &   0.980  \\ 
  0.700 &   1.787 &   1.539 &   1.537 &   1.537 &   1.548 &   1.562 &   0.987 &   0.986 &   0.987  \\ 
  0.750 &   1.818 &   1.562 &   1.561 &   1.560 &   1.572 &   1.584 &   0.987 &   0.986 &   0.986  \\ 
  0.800 &   1.848 &   1.588 &   1.590 &   1.584 &   1.594 &   1.610 &   0.992 &   0.993 &   0.991  \\ 
  0.850 &   1.876 &   1.621 &   1.627 &   1.616 &   1.624 &   1.641 &   0.997 &   1.000 &   0.995  \\ 
  0.900 &   1.903 &   1.665 &   1.676 &   1.658 &   1.667 &   1.683 &   0.998 &   1.005 &   0.996  \\ 
  0.950 &   1.929 &   1.738 &   1.761 &   1.728 &   1.731 &   1.752 &   1.007 &   1.021 &   1.004  \\ 
  0.980 &   1.944 &   1.807 &   1.840 &   1.794 &   1.788 &   1.817 &   1.023 &   1.043 &   1.020  \\ 
\hline \multicolumn{10}{c}{Set  3\,\,\,\,                $X=1$ $\quad$ $Z=0$  $\,\,\,\, \log g = 14.6 \,\,\,\, T_{\rm Edd}= 2.322$ keV  $\quad R=    8.16$ km  $\quad  z=   0.42$} \\ \hline
  0.001 &   0.413 &   1.851 &   1.850 &   1.851 &   1.715 &   1.525 &   0.804 &   0.805 &   0.804  \\ 
  0.003 &   0.543 &   1.703 &   1.703 &   1.702 &   1.595 &   1.438 &   0.876 &   0.876 &   0.876  \\ 
  0.010 &   0.734 &   1.584 &   1.585 &   1.582 &   1.512 &   1.383 &   0.930 &   0.929 &   0.931  \\ 
  0.030 &   0.966 &   1.504 &   1.507 &   1.501 &   1.490 &   1.407 &   0.959 &   0.958 &   0.959  \\ 
  0.050 &   1.098 &   1.474 &   1.478 &   1.471 &   1.479 &   1.425 &   0.967 &   0.967 &   0.967  \\ 
  0.070 &   1.194 &   1.457 &   1.461 &   1.454 &   1.468 &   1.431 &   0.970 &   0.971 &   0.970  \\ 
  0.100 &   1.306 &   1.443 &   1.446 &   1.440 &   1.456 &   1.434 &   0.973 &   0.974 &   0.973  \\ 
  0.150 &   1.445 &   1.431 &   1.433 &   1.429 &   1.446 &   1.435 &   0.975 &   0.976 &   0.975  \\ 
  0.200 &   1.553 &   1.428 &   1.428 &   1.426 &   1.443 &   1.438 &   0.976 &   0.976 &   0.976  \\ 
  0.300 &   1.718 &   1.433 &   1.431 &   1.432 &   1.448 &   1.449 &   0.978 &   0.977 &   0.978  \\ 
  0.400 &   1.847 &   1.449 &   1.445 &   1.448 &   1.462 &   1.468 &   0.981 &   0.978 &   0.980  \\ 
  0.500 &   1.952 &   1.470 &   1.466 &   1.469 &   1.484 &   1.492 &   0.981 &   0.978 &   0.980  \\ 
  0.600 &   2.043 &   1.498 &   1.495 &   1.496 &   1.507 &   1.520 &   0.987 &   0.985 &   0.986  \\ 
  0.700 &   2.124 &   1.535 &   1.534 &   1.533 &   1.544 &   1.556 &   0.987 &   0.986 &   0.987  \\ 
  0.750 &   2.161 &   1.558 &   1.559 &   1.555 &   1.561 &   1.579 &   0.996 &   0.996 &   0.995  \\ 
  0.800 &   2.196 &   1.584 &   1.589 &   1.579 &   1.587 &   1.604 &   0.995 &   0.998 &   0.994  \\ 
  0.850 &   2.229 &   1.619 &   1.627 &   1.613 &   1.619 &   1.637 &   1.000 &   1.005 &   0.999  \\ 
  0.900 &   2.261 &   1.664 &   1.679 &   1.656 &   1.662 &   1.680 &   1.002 &   1.012 &   1.000  \\ 
  0.950 &   2.292 &   1.733 &   1.760 &   1.721 &   1.724 &   1.744 &   1.011 &   1.028 &   1.008  \\ 
  0.980 &   2.310 &   1.810 &   1.852 &   1.794 &   1.798 &   1.816 &   1.014 &   1.039 &   1.010  \\ 
\hline \multicolumn{10}{c}{Set  4\,\,\,\,     $X=0.74$ $\quad$ $Z=Z_{\odot}$  $\,\,\,\, \log g = 14.0 \,\,\,\, T_{\rm Edd}= 1.703$ keV  $\quad R=   14.80$ km  $\quad  z=   0.18$} \\ \hline
  0.001 &   0.303 &   2.157 &   2.148 &   2.165 &   1.187 &   0.976 &   0.096 &   0.097 &   0.096  \\ 
  0.003 &   0.398 &   1.740 &   1.724 &   1.750 &   1.130 &   1.176 &   0.273 &   0.277 &   0.270  \\ 
  0.010 &   0.538 &   1.448 &   1.431 &   1.459 &   1.156 &   1.240 &   0.638 &   0.648 &   0.631  \\ 
  0.030 &   0.709 &   1.368 &   1.365 &   1.366 &   1.266 &   1.203 &   0.890 &   0.891 &   0.891  \\ 
  0.050 &   0.805 &   1.347 &   1.361 &   1.336 &   1.326 &   1.129 &   0.958 &   0.956 &   0.962  \\ 
  0.070 &   0.876 &   1.328 &   1.352 &   1.310 &   1.326 &   1.264 &   0.989 &   0.988 &   0.993  \\ 
  0.100 &   0.957 &   1.310 &   1.340 &   1.290 &   1.309 &   1.257 &   1.000 &   1.003 &   1.004  \\ 
  0.150 &   1.060 &   1.317 &   1.342 &   1.300 &   1.322 &   1.317 &   0.986 &   0.992 &   0.987  \\ 
  0.200 &   1.139 &   1.341 &   1.356 &   1.331 &   1.354 &   1.348 &   0.975 &   0.979 &   0.974  \\ 
  0.300 &   1.260 &   1.392 &   1.395 &   1.390 &   1.405 &   1.412 &   0.979 &   0.981 &   0.979  \\ 
  0.400 &   1.354 &   1.428 &   1.423 &   1.428 &   1.446 &   1.447 &   0.973 &   0.971 &   0.973  \\ 
  0.500 &   1.432 &   1.457 &   1.451 &   1.457 &   1.473 &   1.478 &   0.976 &   0.973 &   0.976  \\ 
  0.600 &   1.498 &   1.484 &   1.480 &   1.483 &   1.497 &   1.508 &   0.982 &   0.979 &   0.981  \\ 
  0.700 &   1.557 &   1.516 &   1.513 &   1.513 &   1.525 &   1.542 &   0.987 &   0.985 &   0.986  \\ 
  0.750 &   1.584 &   1.535 &   1.534 &   1.532 &   1.543 &   1.561 &   0.989 &   0.988 &   0.988  \\ 
  0.800 &   1.610 &   1.555 &   1.556 &   1.552 &   1.562 &   1.582 &   0.990 &   0.991 &   0.990  \\ 
  0.850 &   1.635 &   1.588 &   1.592 &   1.583 &   1.595 &   1.613 &   0.990 &   0.992 &   0.989  \\ 
  0.900 &   1.658 &   1.635 &   1.643 &   1.629 &   1.634 &   1.658 &   1.001 &   1.006 &   1.000  \\ 
  0.950 &   1.681 &   1.700 &   1.716 &   1.692 &   1.699 &   1.720 &   1.002 &   1.011 &   1.000  \\ 
  0.980 &   1.694 &   1.750 &   1.769 &   1.741 &   1.731 &   1.767 &   1.024 &   1.036 &   1.022  \\ 
\hline \multicolumn{10}{c}{Set  5\,\,\,\,     $X=0.74$ $\quad$ $Z=Z_{\odot}$  $\,\,\,\, \log g = 14.3 \,\,\,\, T_{\rm Edd}= 2.024$ keV  $\quad R=   10.88$ km  $\quad  z=   0.27$} \\ \hline
  0.001 &   0.360 &   2.001 &   1.973 &   2.015 &   1.265 &   1.298 &   0.208 &   0.214 &   0.205  \\ 
  0.003 &   0.474 &   1.651 &   1.622 &   1.667 &   1.204 &   1.272 &   0.445 &   0.459 &   0.438  \\ 
  0.010 &   0.640 &   1.427 &   1.409 &   1.433 &   1.221 &   1.073 &   0.771 &   0.781 &   0.767  \\ 
  0.030 &   0.842 &   1.329 &   1.347 &   1.314 &   1.305 &   1.081 &   0.964 &   0.959 &   0.971  \\ 
  0.050 &   0.957 &   1.287 &   1.319 &   1.265 &   1.289 &   1.231 &   1.004 &   1.002 &   1.012  \\ 
  0.070 &   1.041 &   1.269 &   1.304 &   1.248 &   1.270 &   1.269 &   1.006 &   1.006 &   1.010  \\ 
  0.100 &   1.138 &   1.277 &   1.305 &   1.261 &   1.280 &   1.283 &   0.986 &   0.990 &   0.987  \\ 
  0.150 &   1.259 &   1.313 &   1.326 &   1.306 &   1.328 &   1.323 &   0.966 &   0.970 &   0.965  \\ 
  0.200 &   1.353 &   1.351 &   1.353 &   1.351 &   1.365 &   1.368 &   0.976 &   0.976 &   0.976  \\ 
  0.300 &   1.498 &   1.400 &   1.393 &   1.404 &   1.413 &   1.418 &   0.980 &   0.976 &   0.981  \\ 
  0.400 &   1.609 &   1.428 &   1.419 &   1.429 &   1.444 &   1.447 &   0.976 &   0.971 &   0.977  \\ 
  0.500 &   1.702 &   1.452 &   1.445 &   1.451 &   1.465 &   1.477 &   0.981 &   0.977 &   0.980  \\ 
  0.600 &   1.781 &   1.477 &   1.472 &   1.475 &   1.487 &   1.506 &   0.986 &   0.983 &   0.985  \\ 
  0.700 &   1.851 &   1.507 &   1.505 &   1.505 &   1.518 &   1.537 &   0.985 &   0.983 &   0.984  \\ 
  0.750 &   1.883 &   1.529 &   1.528 &   1.526 &   1.536 &   1.558 &   0.991 &   0.990 &   0.990  \\ 
  0.800 &   1.914 &   1.556 &   1.556 &   1.552 &   1.561 &   1.582 &   0.992 &   0.992 &   0.991  \\ 
  0.850 &   1.943 &   1.586 &   1.591 &   1.582 &   1.589 &   1.612 &   0.996 &   0.999 &   0.995  \\ 
  0.900 &   1.971 &   1.629 &   1.638 &   1.622 &   1.629 &   1.649 &   1.000 &   1.005 &   0.998  \\ 
  0.950 &   1.998 &   1.692 &   1.710 &   1.683 &   1.691 &   1.707 &   1.001 &   1.013 &   0.999  \\ 
  0.980 &   2.013 &   1.750 &   1.774 &   1.740 &   1.729 &   1.762 &   1.026 &   1.041 &   1.023  \\ 
\hline \multicolumn{10}{c}{Set  6\,\,\,\,     $X=0.74$ $\quad$ $Z=Z_{\odot}$  $\,\,\,\, \log g = 14.6 \,\,\,\, T_{\rm Edd}= 2.405$ keV  $\quad R=    8.16$ km  $\quad  z=   0.42$} \\ \hline
  0.001 &   0.428 &   1.946 &   1.932 &   1.951 &   1.357 &   1.101 &   0.327 &   0.332 &   0.325  \\ 
  0.003 &   0.563 &   1.636 &   1.627 &   1.634 &   1.281 &   1.203 &   0.572 &   0.577 &   0.573  \\ 
  0.010 &   0.761 &   1.397 &   1.412 &   1.384 &   1.284 &   1.172 &   0.871 &   0.862 &   0.880  \\ 
  0.030 &   1.001 &   1.248 &   1.289 &   1.225 &   1.264 &   1.187 &   1.027 &   1.014 &   1.041  \\ 
  0.050 &   1.137 &   1.223 &   1.260 &   1.202 &   1.227 &   1.250 &   1.014 &   1.010 &   1.024  \\ 
  0.070 &   1.237 &   1.239 &   1.266 &   1.226 &   1.238 &   1.250 &   0.985 &   0.987 &   0.988  \\ 
  0.100 &   1.352 &   1.275 &   1.288 &   1.272 &   1.291 &   1.289 &   0.961 &   0.963 &   0.961  \\ 
  0.150 &   1.497 &   1.328 &   1.325 &   1.333 &   1.354 &   1.331 &   0.953 &   0.952 &   0.953  \\ 
  0.200 &   1.608 &   1.363 &   1.353 &   1.369 &   1.388 &   1.368 &   0.959 &   0.955 &   0.960  \\ 
  0.300 &   1.780 &   1.404 &   1.395 &   1.407 &   1.416 &   1.418 &   0.982 &   0.977 &   0.982  \\ 
  0.400 &   1.913 &   1.424 &   1.418 &   1.424 &   1.436 &   1.447 &   0.982 &   0.978 &   0.982  \\ 
  0.500 &   2.022 &   1.444 &   1.441 &   1.443 &   1.455 &   1.475 &   0.985 &   0.982 &   0.984  \\ 
  0.600 &   2.117 &   1.470 &   1.468 &   1.468 &   1.478 &   1.500 &   0.988 &   0.987 &   0.988  \\ 
  0.700 &   2.200 &   1.504 &   1.503 &   1.501 &   1.511 &   1.531 &   0.990 &   0.990 &   0.990  \\ 
  0.750 &   2.238 &   1.526 &   1.528 &   1.522 &   1.529 &   1.552 &   0.996 &   0.996 &   0.995  \\ 
  0.800 &   2.275 &   1.557 &   1.560 &   1.553 &   1.553 &   1.579 &   1.005 &   1.007 &   1.004  \\ 
  0.850 &   2.309 &   1.584 &   1.591 &   1.579 &   1.587 &   1.603 &   0.996 &   1.000 &   0.995  \\ 
  0.900 &   2.342 &   1.629 &   1.641 &   1.622 &   1.628 &   1.646 &   1.002 &   1.009 &   1.000  \\ 
  0.950 &   2.374 &   1.699 &   1.722 &   1.689 &   1.692 &   1.710 &   1.009 &   1.023 &   1.006  \\ 
  0.980 &   2.393 &   1.770 &   1.806 &   1.756 &   1.758 &   1.775 &   1.014 &   1.037 &   1.010  \\ 
\hline \multicolumn{10}{c}{Set  7\,\,\,\, $X=0.74$  $\quad$ $Z=0.3Z_{\odot}$  $\,\,\,\, \log g = 14.0 \,\,\,\, T_{\rm Edd}= 1.703$ keV  $\quad R=   14.80$ km  $\quad  z=   0.18$} \\ \hline
  0.001 &   0.303 &   2.024 &   2.019 &   2.028 &   1.351 &   0.980 &   0.224 &   0.226 &   0.223  \\ 
  0.003 &   0.398 &   1.742 &   1.731 &   1.749 &   1.271 &   1.180 &   0.426 &   0.431 &   0.423  \\ 
  0.010 &   0.538 &   1.541 &   1.527 &   1.547 &   1.269 &   1.241 &   0.715 &   0.722 &   0.711  \\ 
  0.030 &   0.709 &   1.441 &   1.444 &   1.435 &   1.362 &   1.232 &   0.910 &   0.909 &   0.912  \\ 
  0.050 &   0.805 &   1.402 &   1.415 &   1.390 &   1.386 &   1.297 &   0.960 &   0.958 &   0.964  \\ 
  0.070 &   0.876 &   1.378 &   1.397 &   1.363 &   1.376 &   1.303 &   0.976 &   0.977 &   0.979  \\ 
  0.100 &   0.957 &   1.362 &   1.383 &   1.347 &   1.365 &   1.311 &   0.978 &   0.981 &   0.980  \\ 
  0.150 &   1.060 &   1.370 &   1.384 &   1.359 &   1.379 &   1.362 &   0.975 &   0.979 &   0.975  \\ 
  0.200 &   1.139 &   1.386 &   1.393 &   1.381 &   1.407 &   1.387 &   0.963 &   0.965 &   0.962  \\ 
  0.300 &   1.260 &   1.426 &   1.424 &   1.425 &   1.440 &   1.441 &   0.978 &   0.977 &   0.978  \\ 
  0.400 &   1.354 &   1.449 &   1.444 &   1.448 &   1.464 &   1.467 &   0.975 &   0.973 &   0.975  \\ 
  0.500 &   1.432 &   1.470 &   1.464 &   1.469 &   1.486 &   1.490 &   0.977 &   0.974 &   0.977  \\ 
  0.600 &   1.498 &   1.494 &   1.489 &   1.493 &   1.507 &   1.517 &   0.982 &   0.979 &   0.981  \\ 
  0.700 &   1.557 &   1.522 &   1.519 &   1.520 &   1.537 &   1.546 &   0.979 &   0.977 &   0.979  \\ 
  0.750 &   1.584 &   1.546 &   1.544 &   1.544 &   1.554 &   1.570 &   0.989 &   0.988 &   0.989  \\ 
  0.800 &   1.610 &   1.571 &   1.570 &   1.568 &   1.579 &   1.594 &   0.989 &   0.988 &   0.988  \\ 
  0.850 &   1.635 &   1.605 &   1.607 &   1.601 &   1.609 &   1.627 &   0.994 &   0.995 &   0.993  \\ 
  0.900 &   1.658 &   1.647 &   1.655 &   1.642 &   1.650 &   1.668 &   0.996 &   1.001 &   0.995  \\ 
  0.950 &   1.681 &   1.720 &   1.737 &   1.711 &   1.716 &   1.737 &   1.004 &   1.014 &   1.002  \\ 
  0.980 &   1.694 &   1.785 &   1.817 &   1.772 &   1.797 &   1.799 &   0.986 &   1.004 &   0.983  \\ 
\hline \multicolumn{10}{c}{Set  8\,\,\,\, $X=0.74$  $\quad$ $Z=0.3Z_{\odot}$  $\,\,\,\, \log g = 14.3 \,\,\,\, T_{\rm Edd}= 2.024$ keV  $\quad R=   10.88$ km  $\quad  z=   0.27$} \\ \hline
  0.001 &   0.360 &   1.925 &   1.909 &   1.933 &   1.406 &   1.298 &   0.364 &   0.371 &   0.360  \\ 
  0.003 &   0.474 &   1.685 &   1.669 &   1.692 &   1.329 &   1.269 &   0.578 &   0.587 &   0.574  \\ 
  0.010 &   0.640 &   1.505 &   1.501 &   1.504 &   1.326 &   1.087 &   0.823 &   0.825 &   0.823  \\ 
  0.030 &   0.842 &   1.386 &   1.404 &   1.373 &   1.364 &   1.281 &   0.963 &   0.960 &   0.968  \\ 
  0.050 &   0.957 &   1.344 &   1.368 &   1.329 &   1.340 &   1.252 &   0.982 &   0.982 &   0.986  \\ 
  0.070 &   1.041 &   1.330 &   1.352 &   1.317 &   1.332 &   1.304 &   0.977 &   0.980 &   0.980  \\ 
  0.100 &   1.138 &   1.337 &   1.352 &   1.328 &   1.348 &   1.322 &   0.967 &   0.970 &   0.967  \\ 
  0.150 &   1.259 &   1.364 &   1.368 &   1.363 &   1.386 &   1.364 &   0.959 &   0.960 &   0.959  \\ 
  0.200 &   1.353 &   1.392 &   1.389 &   1.393 &   1.410 &   1.399 &   0.972 &   0.970 &   0.972  \\ 
  0.300 &   1.498 &   1.420 &   1.413 &   1.422 &   1.438 &   1.434 &   0.973 &   0.969 &   0.973  \\ 
  0.400 &   1.609 &   1.440 &   1.433 &   1.441 &   1.456 &   1.459 &   0.978 &   0.973 &   0.977  \\ 
  0.500 &   1.702 &   1.460 &   1.453 &   1.460 &   1.473 &   1.484 &   0.981 &   0.977 &   0.981  \\ 
  0.600 &   1.781 &   1.485 &   1.479 &   1.484 &   1.496 &   1.510 &   0.984 &   0.980 &   0.984  \\ 
  0.700 &   1.851 &   1.520 &   1.515 &   1.518 &   1.524 &   1.545 &   0.989 &   0.986 &   0.989  \\ 
  0.750 &   1.883 &   1.537 &   1.535 &   1.535 &   1.546 &   1.562 &   0.988 &   0.986 &   0.987  \\ 
  0.800 &   1.914 &   1.564 &   1.563 &   1.561 &   1.571 &   1.587 &   0.991 &   0.991 &   0.990  \\ 
  0.850 &   1.943 &   1.598 &   1.602 &   1.594 &   1.603 &   1.620 &   0.994 &   0.996 &   0.993  \\ 
  0.900 &   1.971 &   1.645 &   1.655 &   1.639 &   1.645 &   1.664 &   1.001 &   1.007 &   0.999  \\ 
  0.950 &   1.998 &   1.718 &   1.739 &   1.709 &   1.712 &   1.732 &   1.008 &   1.020 &   1.005  \\ 
  0.980 &   2.013 &   1.776 &   1.800 &   1.765 &   1.746 &   1.786 &   1.036 &   1.052 &   1.033  \\ 
\hline \multicolumn{10}{c}{Set  9\,\,\,\, $X=0.74$  $\quad$ $Z=0.3Z_{\odot}$  $\,\,\,\, \log g = 14.6 \,\,\,\, T_{\rm Edd}= 2.405$ keV  $\quad R=    8.16$ km  $\quad  z=   0.42$} \\ \hline
  0.001 &   0.428 &   1.882 &   1.876 &   1.883 &   1.476 &   1.106 &   0.486 &   0.489 &   0.485  \\ 
  0.003 &   0.563 &   1.651 &   1.652 &   1.647 &   1.391 &   1.206 &   0.694 &   0.693 &   0.697  \\ 
  0.010 &   0.761 &   1.449 &   1.467 &   1.436 &   1.366 &   1.178 &   0.910 &   0.900 &   0.919  \\ 
  0.030 &   1.001 &   1.317 &   1.344 &   1.302 &   1.312 &   1.209 &   0.992 &   0.986 &   0.998  \\ 
  0.050 &   1.137 &   1.298 &   1.319 &   1.288 &   1.294 &   1.285 &   0.977 &   0.976 &   0.980  \\ 
  0.070 &   1.237 &   1.310 &   1.323 &   1.306 &   1.315 &   1.301 &   0.961 &   0.963 &   0.962  \\ 
  0.100 &   1.352 &   1.338 &   1.341 &   1.340 &   1.358 &   1.332 &   0.954 &   0.954 &   0.954  \\ 
  0.150 &   1.497 &   1.372 &   1.366 &   1.376 &   1.398 &   1.370 &   0.956 &   0.954 &   0.956  \\ 
  0.200 &   1.608 &   1.394 &   1.386 &   1.397 &   1.415 &   1.398 &   0.968 &   0.964 &   0.968  \\ 
  0.300 &   1.780 &   1.415 &   1.409 &   1.416 &   1.430 &   1.430 &   0.978 &   0.975 &   0.978  \\ 
  0.400 &   1.913 &   1.431 &   1.424 &   1.430 &   1.446 &   1.451 &   0.978 &   0.975 &   0.978  \\ 
  0.500 &   2.022 &   1.456 &   1.450 &   1.455 &   1.460 &   1.481 &   0.994 &   0.990 &   0.993  \\ 
  0.600 &   2.117 &   1.475 &   1.471 &   1.474 &   1.489 &   1.501 &   0.981 &   0.978 &   0.981  \\ 
  0.700 &   2.200 &   1.514 &   1.511 &   1.513 &   1.523 &   1.537 &   0.988 &   0.986 &   0.987  \\ 
  0.750 &   2.238 &   1.535 &   1.535 &   1.533 &   1.541 &   1.558 &   0.992 &   0.991 &   0.991  \\ 
  0.800 &   2.275 &   1.566 &   1.567 &   1.562 &   1.568 &   1.586 &   0.996 &   0.997 &   0.995  \\ 
  0.850 &   2.309 &   1.596 &   1.602 &   1.592 &   1.598 &   1.614 &   0.997 &   1.001 &   0.996  \\ 
  0.900 &   2.342 &   1.641 &   1.654 &   1.633 &   1.639 &   1.657 &   1.002 &   1.011 &   1.000  \\ 
  0.950 &   2.374 &   1.714 &   1.740 &   1.703 &   1.707 &   1.724 &   1.009 &   1.026 &   1.006  \\ 
  0.980 &   2.393 &   1.808 &   1.851 &   1.792 &   1.790 &   1.811 &   1.021 &   1.048 &   1.017  \\ 
\hline \multicolumn{10}{c}{Set 10\,\,\,\, $X=0.74$  $\quad$ $Z=0.1Z_{\odot}$  $\,\,\,\, \log g = 14.0 \,\,\,\, T_{\rm Edd}= 1.703$ keV  $\quad R=   14.80$ km  $\quad  z=   0.18$} \\ \hline
  0.001 &   0.303 &   1.975 &   1.969 &   1.979 &   1.504 &   1.360 &   0.387 &   0.390 &   0.386  \\ 
  0.003 &   0.398 &   1.767 &   1.758 &   1.772 &   1.406 &   1.188 &   0.572 &   0.577 &   0.569  \\ 
  0.010 &   0.538 &   1.604 &   1.597 &   1.606 &   1.378 &   1.255 &   0.788 &   0.791 &   0.787  \\ 
  0.030 &   0.709 &   1.495 &   1.500 &   1.490 &   1.437 &   1.261 &   0.924 &   0.923 &   0.926  \\ 
  0.050 &   0.805 &   1.452 &   1.462 &   1.445 &   1.442 &   1.341 &   0.956 &   0.956 &   0.958  \\ 
  0.070 &   0.876 &   1.430 &   1.442 &   1.421 &   1.432 &   1.345 &   0.966 &   0.966 &   0.967  \\ 
  0.100 &   0.957 &   1.414 &   1.426 &   1.406 &   1.424 &   1.376 &   0.967 &   0.969 &   0.968  \\ 
  0.150 &   1.060 &   1.412 &   1.419 &   1.406 &   1.430 &   1.400 &   0.965 &   0.967 &   0.965  \\ 
  0.200 &   1.139 &   1.421 &   1.424 &   1.418 &   1.441 &   1.422 &   0.966 &   0.967 &   0.965  \\ 
  0.300 &   1.260 &   1.440 &   1.438 &   1.439 &   1.459 &   1.453 &   0.971 &   0.970 &   0.971  \\ 
  0.400 &   1.354 &   1.457 &   1.452 &   1.456 &   1.474 &   1.474 &   0.975 &   0.972 &   0.975  \\ 
  0.500 &   1.432 &   1.474 &   1.468 &   1.473 &   1.489 &   1.493 &   0.978 &   0.975 &   0.978  \\ 
  0.600 &   1.498 &   1.497 &   1.492 &   1.496 &   1.510 &   1.519 &   0.982 &   0.979 &   0.981  \\ 
  0.700 &   1.557 &   1.522 &   1.519 &   1.520 &   1.539 &   1.546 &   0.977 &   0.975 &   0.977  \\ 
  0.750 &   1.584 &   1.551 &   1.548 &   1.550 &   1.558 &   1.574 &   0.990 &   0.988 &   0.990  \\ 
  0.800 &   1.610 &   1.574 &   1.574 &   1.571 &   1.580 &   1.597 &   0.992 &   0.992 &   0.991  \\ 
  0.850 &   1.635 &   1.608 &   1.610 &   1.604 &   1.613 &   1.629 &   0.992 &   0.993 &   0.991  \\ 
  0.900 &   1.658 &   1.651 &   1.659 &   1.646 &   1.655 &   1.672 &   0.995 &   0.999 &   0.994  \\ 
  0.950 &   1.681 &   1.730 &   1.748 &   1.721 &   1.724 &   1.746 &   1.006 &   1.017 &   1.004  \\ 
  0.980 &   1.694 &   1.787 &   1.811 &   1.778 &   1.763 &   1.802 &   1.031 &   1.044 &   1.028  \\ 
\hline \multicolumn{10}{c}{Set 11\,\,\,\, $X=0.74$  $\quad$ $Z=0.1Z_{\odot}$  $\,\,\,\, \log g = 14.3 \,\,\,\, T_{\rm Edd}= 2.024$ keV  $\quad R=   10.88$ km  $\quad  z=   0.27$} \\ \hline
  0.001 &   0.360 &   1.910 &   1.900 &   1.915 &   1.537 &   1.302 &   0.519 &   0.525 &   0.516  \\ 
  0.003 &   0.474 &   1.719 &   1.711 &   1.722 &   1.445 &   1.283 &   0.693 &   0.698 &   0.691  \\ 
  0.010 &   0.640 &   1.560 &   1.560 &   1.556 &   1.416 &   1.275 &   0.866 &   0.866 &   0.868  \\ 
  0.030 &   0.842 &   1.444 &   1.456 &   1.435 &   1.424 &   1.314 &   0.956 &   0.955 &   0.959  \\ 
  0.050 &   0.957 &   1.407 &   1.420 &   1.398 &   1.404 &   1.319 &   0.967 &   0.968 &   0.969  \\ 
  0.070 &   1.041 &   1.393 &   1.404 &   1.386 &   1.400 &   1.354 &   0.966 &   0.967 &   0.966  \\ 
  0.100 &   1.138 &   1.390 &   1.397 &   1.387 &   1.407 &   1.369 &   0.962 &   0.964 &   0.962  \\ 
  0.150 &   1.259 &   1.402 &   1.402 &   1.401 &   1.424 &   1.399 &   0.962 &   0.962 &   0.962  \\ 
  0.200 &   1.353 &   1.414 &   1.410 &   1.414 &   1.435 &   1.419 &   0.967 &   0.965 &   0.967  \\ 
  0.300 &   1.498 &   1.430 &   1.424 &   1.430 &   1.448 &   1.444 &   0.973 &   0.970 &   0.973  \\ 
  0.400 &   1.609 &   1.444 &   1.437 &   1.444 &   1.460 &   1.463 &   0.977 &   0.973 &   0.977  \\ 
  0.500 &   1.702 &   1.463 &   1.455 &   1.462 &   1.476 &   1.485 &   0.981 &   0.976 &   0.981  \\ 
  0.600 &   1.781 &   1.488 &   1.481 &   1.487 &   1.499 &   1.512 &   0.984 &   0.980 &   0.984  \\ 
  0.700 &   1.851 &   1.522 &   1.517 &   1.521 &   1.532 &   1.546 &   0.987 &   0.983 &   0.987  \\ 
  0.750 &   1.883 &   1.540 &   1.537 &   1.539 &   1.553 &   1.563 &   0.983 &   0.981 &   0.983  \\ 
  0.800 &   1.914 &   1.569 &   1.569 &   1.566 &   1.574 &   1.592 &   0.994 &   0.994 &   0.993  \\ 
  0.850 &   1.943 &   1.603 &   1.607 &   1.599 &   1.607 &   1.625 &   0.995 &   0.997 &   0.994  \\ 
  0.900 &   1.971 &   1.649 &   1.659 &   1.643 &   1.650 &   1.667 &   0.999 &   1.005 &   0.997  \\ 
  0.950 &   1.998 &   1.728 &   1.750 &   1.718 &   1.724 &   1.741 &   1.006 &   1.019 &   1.003  \\ 
  0.980 &   2.013 &   1.789 &   1.816 &   1.778 &   1.758 &   1.799 &   1.038 &   1.055 &   1.035  \\ 
\hline \multicolumn{10}{c}{Set 12\,\,\,\, $X=0.74$  $\quad$ $Z=0.1Z_{\odot}$  $\,\,\,\, \log g = 14.6 \,\,\,\, T_{\rm Edd}= 2.405$ keV  $\quad R=    8.16$ km  $\quad  z=   0.42$} \\ \hline
  0.001 &   0.428 &   1.868 &   1.866 &   1.868 &   1.581 &   1.377 &   0.623 &   0.625 &   0.623  \\ 
  0.003 &   0.563 &   1.675 &   1.679 &   1.673 &   1.485 &   1.217 &   0.784 &   0.782 &   0.786  \\ 
  0.010 &   0.761 &   1.503 &   1.517 &   1.495 &   1.431 &   1.193 &   0.925 &   0.919 &   0.930  \\ 
  0.030 &   1.001 &   1.394 &   1.408 &   1.386 &   1.381 &   1.330 &   0.969 &   0.968 &   0.972  \\ 
  0.050 &   1.137 &   1.373 &   1.383 &   1.369 &   1.374 &   1.331 &   0.964 &   0.964 &   0.964  \\ 
  0.070 &   1.237 &   1.374 &   1.379 &   1.374 &   1.386 &   1.351 &   0.959 &   0.960 &   0.959  \\ 
  0.100 &   1.352 &   1.385 &   1.384 &   1.386 &   1.405 &   1.373 &   0.959 &   0.959 &   0.959  \\ 
  0.150 &   1.497 &   1.399 &   1.395 &   1.400 &   1.420 &   1.399 &   0.965 &   0.963 &   0.965  \\ 
  0.200 &   1.608 &   1.408 &   1.403 &   1.409 &   1.427 &   1.414 &   0.972 &   0.970 &   0.972  \\ 
  0.300 &   1.780 &   1.419 &   1.414 &   1.419 &   1.434 &   1.435 &   0.977 &   0.975 &   0.977  \\ 
  0.400 &   1.913 &   1.433 &   1.427 &   1.433 &   1.448 &   1.453 &   0.980 &   0.976 &   0.979  \\ 
  0.500 &   2.022 &   1.454 &   1.448 &   1.453 &   1.466 &   1.477 &   0.983 &   0.979 &   0.983  \\ 
  0.600 &   2.117 &   1.478 &   1.473 &   1.477 &   1.491 &   1.502 &   0.982 &   0.979 &   0.982  \\ 
  0.700 &   2.200 &   1.516 &   1.513 &   1.515 &   1.524 &   1.539 &   0.988 &   0.986 &   0.988  \\ 
  0.750 &   2.238 &   1.537 &   1.537 &   1.534 &   1.543 &   1.559 &   0.992 &   0.991 &   0.991  \\ 
  0.800 &   2.275 &   1.567 &   1.569 &   1.564 &   1.569 &   1.588 &   0.997 &   0.998 &   0.996  \\ 
  0.850 &   2.309 &   1.600 &   1.605 &   1.595 &   1.602 &   1.617 &   0.997 &   1.000 &   0.996  \\ 
  0.900 &   2.342 &   1.646 &   1.659 &   1.639 &   1.647 &   1.662 &   0.999 &   1.007 &   0.997  \\ 
  0.950 &   2.374 &   1.718 &   1.745 &   1.707 &   1.712 &   1.728 &   1.008 &   1.025 &   1.005  \\ 
  0.980 &   2.393 &   1.795 &   1.839 &   1.778 &   1.798 &   1.798 &   0.997 &   1.024 &   0.992  \\ 
\hline \multicolumn{10}{c}{Set 13\,\,\,\,$X=0.74$  $\quad$ $Z=0.01Z_{\odot}$  $\,\,\,\, \log g = 14.0 \,\,\,\, T_{\rm Edd}= 1.703$ keV  $\quad R=   14.80$ km  $\quad  z=   0.18$} \\ \hline
  0.001 &   0.303 &   1.949 &   1.945 &   1.951 &   1.713 &   1.521 &   0.668 &   0.672 &   0.666  \\ 
  0.003 &   0.398 &   1.780 &   1.777 &   1.782 &   1.591 &   1.427 &   0.780 &   0.783 &   0.779  \\ 
  0.010 &   0.538 &   1.644 &   1.642 &   1.644 &   1.512 &   1.305 &   0.880 &   0.881 &   0.880  \\ 
  0.030 &   0.709 &   1.550 &   1.551 &   1.548 &   1.512 &   1.361 &   0.938 &   0.937 &   0.938  \\ 
  0.050 &   0.805 &   1.515 &   1.517 &   1.512 &   1.512 &   1.398 &   0.953 &   0.953 &   0.954  \\ 
  0.070 &   0.876 &   1.495 &   1.497 &   1.492 &   1.503 &   1.412 &   0.960 &   0.960 &   0.960  \\ 
  0.100 &   0.957 &   1.477 &   1.480 &   1.474 &   1.492 &   1.438 &   0.965 &   0.966 &   0.965  \\ 
  0.150 &   1.060 &   1.462 &   1.463 &   1.459 &   1.480 &   1.449 &   0.969 &   0.969 &   0.969  \\ 
  0.200 &   1.139 &   1.456 &   1.456 &   1.453 &   1.474 &   1.456 &   0.971 &   0.971 &   0.971  \\ 
  0.300 &   1.260 &   1.455 &   1.452 &   1.453 &   1.472 &   1.465 &   0.974 &   0.973 &   0.974  \\ 
  0.400 &   1.354 &   1.462 &   1.458 &   1.460 &   1.478 &   1.479 &   0.977 &   0.974 &   0.976  \\ 
  0.500 &   1.432 &   1.476 &   1.471 &   1.475 &   1.490 &   1.495 &   0.979 &   0.976 &   0.978  \\ 
  0.600 &   1.498 &   1.497 &   1.492 &   1.496 &   1.511 &   1.520 &   0.981 &   0.978 &   0.981  \\ 
  0.700 &   1.557 &   1.529 &   1.524 &   1.528 &   1.542 &   1.552 &   0.982 &   0.979 &   0.982  \\ 
  0.750 &   1.584 &   1.549 &   1.547 &   1.547 &   1.560 &   1.572 &   0.986 &   0.984 &   0.985  \\ 
  0.800 &   1.610 &   1.575 &   1.574 &   1.573 &   1.586 &   1.597 &   0.986 &   0.985 &   0.986  \\ 
  0.850 &   1.635 &   1.608 &   1.611 &   1.604 &   1.613 &   1.630 &   0.993 &   0.994 &   0.992  \\ 
  0.900 &   1.658 &   1.654 &   1.661 &   1.649 &   1.659 &   1.674 &   0.994 &   0.998 &   0.993  \\ 
  0.950 &   1.681 &   1.732 &   1.750 &   1.724 &   1.730 &   1.748 &   1.004 &   1.014 &   1.002  \\ 
  0.980 &   1.694 &   1.803 &   1.836 &   1.789 &   1.815 &   1.814 &   0.986 &   1.006 &   0.983  \\ 
\hline \multicolumn{10}{c}{Set 14\,\,\,\,$X=0.74$  $\quad$ $Z=0.01Z_{\odot}$  $\,\,\,\, \log g = 14.3 \,\,\,\, T_{\rm Edd}= 2.024$ keV  $\quad R=   10.88$ km  $\quad  z=   0.27$} \\ \hline
  0.001 &   0.360 &   1.900 &   1.896 &   1.902 &   1.700 &   1.520 &   0.733 &   0.737 &   0.731  \\ 
  0.003 &   0.474 &   1.739 &   1.735 &   1.740 &   1.580 &   1.423 &   0.830 &   0.832 &   0.829  \\ 
  0.010 &   0.640 &   1.606 &   1.606 &   1.606 &   1.509 &   1.348 &   0.909 &   0.909 &   0.909  \\ 
  0.030 &   0.842 &   1.516 &   1.518 &   1.513 &   1.502 &   1.378 &   0.951 &   0.951 &   0.951  \\ 
  0.050 &   0.957 &   1.483 &   1.485 &   1.481 &   1.491 &   1.406 &   0.961 &   0.961 &   0.961  \\ 
  0.070 &   1.041 &   1.466 &   1.468 &   1.463 &   1.479 &   1.418 &   0.965 &   0.966 &   0.965  \\ 
  0.100 &   1.138 &   1.451 &   1.452 &   1.449 &   1.468 &   1.429 &   0.968 &   0.968 &   0.968  \\ 
  0.150 &   1.259 &   1.440 &   1.439 &   1.438 &   1.457 &   1.436 &   0.971 &   0.971 &   0.971  \\ 
  0.200 &   1.353 &   1.435 &   1.433 &   1.434 &   1.453 &   1.441 &   0.974 &   0.973 &   0.973  \\ 
  0.300 &   1.498 &   1.437 &   1.432 &   1.436 &   1.452 &   1.451 &   0.977 &   0.974 &   0.976  \\ 
  0.400 &   1.609 &   1.446 &   1.440 &   1.446 &   1.462 &   1.465 &   0.978 &   0.974 &   0.978  \\ 
  0.500 &   1.702 &   1.459 &   1.452 &   1.458 &   1.479 &   1.482 &   0.971 &   0.967 &   0.971  \\ 
  0.600 &   1.781 &   1.494 &   1.486 &   1.493 &   1.498 &   1.517 &   0.994 &   0.990 &   0.994  \\ 
  0.700 &   1.851 &   1.525 &   1.518 &   1.524 &   1.534 &   1.548 &   0.988 &   0.983 &   0.987  \\ 
  0.750 &   1.883 &   1.546 &   1.542 &   1.545 &   1.555 &   1.568 &   0.988 &   0.985 &   0.988  \\ 
  0.800 &   1.914 &   1.571 &   1.570 &   1.569 &   1.579 &   1.593 &   0.990 &   0.989 &   0.990  \\ 
  0.850 &   1.943 &   1.604 &   1.607 &   1.600 &   1.609 &   1.624 &   0.993 &   0.995 &   0.992  \\ 
  0.900 &   1.971 &   1.651 &   1.661 &   1.645 &   1.652 &   1.669 &   0.999 &   1.005 &   0.997  \\ 
  0.950 &   1.998 &   1.728 &   1.749 &   1.718 &   1.721 &   1.741 &   1.008 &   1.021 &   1.006  \\ 
  0.980 &   2.013 &   1.775 &   1.800 &   1.765 &   1.746 &   1.786 &   1.037 &   1.052 &   1.034  \\ 
\hline \multicolumn{10}{c}{Set 15\,\,\,\,$X=0.74$  $\quad$ $Z=0.01Z_{\odot}$  $\,\,\,\, \log g = 14.6 \,\,\,\, T_{\rm Edd}= 2.405$ keV  $\quad R=    8.16$ km  $\quad  z=   0.42$} \\ \hline
  0.001 &   0.428 &   1.861 &   1.858 &   1.861 &   1.695 &   1.427 &   0.778 &   0.780 &   0.777  \\ 
  0.003 &   0.563 &   1.702 &   1.702 &   1.701 &   1.576 &   1.427 &   0.864 &   0.864 &   0.864  \\ 
  0.010 &   0.761 &   1.571 &   1.573 &   1.569 &   1.497 &   1.378 &   0.929 &   0.928 &   0.930  \\ 
  0.030 &   1.001 &   1.485 &   1.488 &   1.482 &   1.476 &   1.387 &   0.959 &   0.959 &   0.959  \\ 
  0.050 &   1.137 &   1.456 &   1.459 &   1.454 &   1.465 &   1.402 &   0.965 &   0.965 &   0.965  \\ 
  0.070 &   1.237 &   1.442 &   1.443 &   1.440 &   1.455 &   1.412 &   0.968 &   0.968 &   0.968  \\ 
  0.100 &   1.352 &   1.430 &   1.430 &   1.428 &   1.445 &   1.418 &   0.971 &   0.971 &   0.971  \\ 
  0.150 &   1.497 &   1.421 &   1.420 &   1.419 &   1.437 &   1.422 &   0.974 &   0.974 &   0.974  \\ 
  0.200 &   1.608 &   1.418 &   1.416 &   1.416 &   1.433 &   1.427 &   0.976 &   0.975 &   0.976  \\ 
  0.300 &   1.780 &   1.421 &   1.417 &   1.420 &   1.436 &   1.438 &   0.978 &   0.976 &   0.978  \\ 
  0.400 &   1.913 &   1.434 &   1.429 &   1.434 &   1.448 &   1.454 &   0.980 &   0.976 &   0.980  \\ 
  0.500 &   2.022 &   1.455 &   1.448 &   1.455 &   1.467 &   1.477 &   0.983 &   0.979 &   0.983  \\ 
  0.600 &   2.117 &   1.484 &   1.478 &   1.484 &   1.493 &   1.507 &   0.988 &   0.983 &   0.988  \\ 
  0.700 &   2.200 &   1.518 &   1.514 &   1.516 &   1.526 &   1.540 &   0.989 &   0.987 &   0.989  \\ 
  0.750 &   2.238 &   1.540 &   1.538 &   1.538 &   1.548 &   1.562 &   0.989 &   0.988 &   0.988  \\ 
  0.800 &   2.275 &   1.570 &   1.571 &   1.567 &   1.573 &   1.590 &   0.996 &   0.996 &   0.995  \\ 
  0.850 &   2.309 &   1.601 &   1.607 &   1.596 &   1.603 &   1.619 &   0.998 &   1.002 &   0.997  \\ 
  0.900 &   2.342 &   1.649 &   1.663 &   1.642 &   1.647 &   1.665 &   1.003 &   1.011 &   1.001  \\ 
  0.950 &   2.374 &   1.724 &   1.751 &   1.713 &   1.717 &   1.734 &   1.009 &   1.025 &   1.006  \\ 
  0.980 &   2.393 &   1.785 &   1.816 &   1.773 &   1.750 &   1.791 &   1.042 &   1.062 &   1.039  \\ 
\hline \multicolumn{10}{c}{Set 16\,\,\,\,               $Y=1$  $\quad$ $Z=0$  $\,\,\,\, \log g = 14.0 \,\,\,\, T_{\rm Edd}= 1.955$ keV  $\quad R=   14.80$ km  $\quad  z=   0.18$} \\ \hline
  0.001 &   0.348 &   1.967 &   1.956 &   1.973 &   1.781 &   1.609 &   0.793 &   0.801 &   0.788  \\ 
  0.003 &   0.457 &   1.785 &   1.774 &   1.791 &   1.637 &   1.517 &   0.862 &   0.869 &   0.858  \\ 
  0.010 &   0.618 &   1.641 &   1.628 &   1.646 &   1.552 &   1.423 &   0.914 &   0.917 &   0.912  \\ 
  0.030 &   0.814 &   1.545 &   1.534 &   1.550 &   1.550 &   1.375 &   0.946 &   0.945 &   0.945  \\ 
  0.050 &   0.924 &   1.509 &   1.498 &   1.513 &   1.529 &   1.379 &   0.957 &   0.954 &   0.957  \\ 
  0.070 &   1.005 &   1.488 &   1.477 &   1.491 &   1.508 &   1.390 &   0.962 &   0.959 &   0.962  \\ 
  0.100 &   1.099 &   1.467 &   1.457 &   1.470 &   1.487 &   1.405 &   0.967 &   0.964 &   0.968  \\ 
  0.150 &   1.217 &   1.446 &   1.435 &   1.448 &   1.464 &   1.419 &   0.972 &   0.967 &   0.972  \\ 
  0.200 &   1.307 &   1.433 &   1.421 &   1.435 &   1.450 &   1.424 &   0.974 &   0.968 &   0.974  \\ 
  0.300 &   1.447 &   1.419 &   1.405 &   1.421 &   1.433 &   1.428 &   0.976 &   0.968 &   0.977  \\ 
  0.400 &   1.555 &   1.415 &   1.400 &   1.417 &   1.430 &   1.432 &   0.977 &   0.968 &   0.977  \\ 
  0.500 &   1.644 &   1.419 &   1.402 &   1.421 &   1.434 &   1.441 &   0.978 &   0.967 &   0.978  \\ 
  0.600 &   1.720 &   1.430 &   1.412 &   1.433 &   1.445 &   1.455 &   0.979 &   0.968 &   0.980  \\ 
  0.700 &   1.788 &   1.451 &   1.433 &   1.454 &   1.465 &   1.478 &   0.980 &   0.968 &   0.981  \\ 
  0.750 &   1.819 &   1.468 &   1.451 &   1.470 &   1.480 &   1.494 &   0.983 &   0.971 &   0.983  \\ 
  0.800 &   1.849 &   1.487 &   1.472 &   1.489 &   1.500 &   1.514 &   0.982 &   0.972 &   0.983  \\ 
  0.850 &   1.877 &   1.518 &   1.504 &   1.520 &   1.529 &   1.543 &   0.985 &   0.976 &   0.985  \\ 
  0.900 &   1.904 &   1.556 &   1.548 &   1.556 &   1.569 &   1.579 &   0.984 &   0.978 &   0.983  \\ 
  0.950 &   1.930 &   1.629 &   1.634 &   1.624 &   1.632 &   1.649 &   0.996 &   0.999 &   0.995  \\ 
  0.980 &   1.945 &   1.686 &   1.688 &   1.682 &   1.659 &   1.703 &   1.033 &   1.034 &   1.032  \\ 
\hline \multicolumn{10}{c}{Set 17\,\,\,\,               $Y=1$  $\quad$ $Z=0$  $\,\,\,\, \log g = 14.3 \,\,\,\, T_{\rm Edd}= 2.323$ keV  $\quad R=   10.88$ km  $\quad  z=   0.27$} \\ \hline
  0.001 &   0.413 &   1.924 &   1.913 &   1.929 &   1.751 &   1.598 &   0.821 &   0.828 &   0.817  \\ 
  0.003 &   0.544 &   1.748 &   1.736 &   1.754 &   1.617 &   1.508 &   0.884 &   0.890 &   0.881  \\ 
  0.010 &   0.735 &   1.609 &   1.597 &   1.614 &   1.555 &   1.418 &   0.929 &   0.931 &   0.928  \\ 
  0.030 &   0.967 &   1.517 &   1.505 &   1.521 &   1.530 &   1.378 &   0.956 &   0.954 &   0.955  \\ 
  0.050 &   1.099 &   1.482 &   1.470 &   1.485 &   1.503 &   1.384 &   0.964 &   0.961 &   0.964  \\ 
  0.070 &   1.195 &   1.461 &   1.450 &   1.463 &   1.480 &   1.393 &   0.969 &   0.965 &   0.969  \\ 
  0.100 &   1.306 &   1.440 &   1.429 &   1.443 &   1.458 &   1.403 &   0.973 &   0.968 &   0.973  \\ 
  0.150 &   1.446 &   1.420 &   1.408 &   1.422 &   1.436 &   1.409 &   0.976 &   0.970 &   0.976  \\ 
  0.200 &   1.554 &   1.408 &   1.394 &   1.410 &   1.423 &   1.411 &   0.977 &   0.970 &   0.977  \\ 
  0.300 &   1.719 &   1.396 &   1.380 &   1.399 &   1.411 &   1.412 &   0.978 &   0.968 &   0.979  \\ 
  0.400 &   1.848 &   1.395 &   1.377 &   1.399 &   1.410 &   1.417 &   0.979 &   0.967 &   0.979  \\ 
  0.500 &   1.954 &   1.403 &   1.382 &   1.406 &   1.416 &   1.428 &   0.980 &   0.966 &   0.980  \\ 
  0.600 &   2.045 &   1.417 &   1.396 &   1.421 &   1.431 &   1.444 &   0.980 &   0.966 &   0.981  \\ 
  0.700 &   2.125 &   1.443 &   1.423 &   1.446 &   1.452 &   1.471 &   0.987 &   0.972 &   0.988  \\ 
  0.750 &   2.162 &   1.461 &   1.442 &   1.464 &   1.466 &   1.489 &   0.993 &   0.979 &   0.993  \\ 
  0.800 &   2.197 &   1.480 &   1.462 &   1.483 &   1.497 &   1.506 &   0.978 &   0.965 &   0.979  \\ 
  0.850 &   2.231 &   1.514 &   1.500 &   1.515 &   1.521 &   1.539 &   0.989 &   0.979 &   0.989  \\ 
  0.900 &   2.263 &   1.559 &   1.551 &   1.558 &   1.562 &   1.580 &   0.995 &   0.990 &   0.995  \\ 
  0.950 &   2.294 &   1.630 &   1.636 &   1.625 &   1.630 &   1.646 &   1.001 &   1.005 &   0.999  \\ 
  0.980 &   2.312 &   1.690 &   1.699 &   1.683 &   1.664 &   1.702 &   1.032 &   1.039 &   1.030  \\ 
\hline \multicolumn{10}{c}{Set 18\,\,\,\,               $Y=1$  $\quad$ $Z=0$  $\,\,\,\, \log g = 14.6 \,\,\,\, T_{\rm Edd}= 2.761$ keV  $\quad R=    8.16$ km  $\quad  z=   0.42$} \\ \hline
  0.001 &   0.491 &   1.893 &   1.884 &   1.897 &   1.734 &   1.590 &   0.838 &   0.844 &   0.834  \\ 
  0.003 &   0.646 &   1.721 &   1.712 &   1.726 &   1.605 &   1.501 &   0.897 &   0.902 &   0.895  \\ 
  0.010 &   0.873 &   1.584 &   1.575 &   1.588 &   1.546 &   1.412 &   0.939 &   0.940 &   0.938  \\ 
  0.030 &   1.149 &   1.492 &   1.484 &   1.495 &   1.506 &   1.380 &   0.963 &   0.961 &   0.963  \\ 
  0.050 &   1.306 &   1.457 &   1.449 &   1.459 &   1.473 &   1.386 &   0.970 &   0.967 &   0.970  \\ 
  0.070 &   1.420 &   1.437 &   1.429 &   1.438 &   1.452 &   1.392 &   0.973 &   0.971 &   0.974  \\ 
  0.100 &   1.553 &   1.417 &   1.408 &   1.418 &   1.432 &   1.396 &   0.976 &   0.973 &   0.976  \\ 
  0.150 &   1.718 &   1.397 &   1.388 &   1.399 &   1.411 &   1.396 &   0.978 &   0.973 &   0.979  \\ 
  0.200 &   1.847 &   1.387 &   1.375 &   1.388 &   1.400 &   1.394 &   0.979 &   0.973 &   0.979  \\ 
  0.300 &   2.043 &   1.378 &   1.363 &   1.380 &   1.392 &   1.396 &   0.980 &   0.970 &   0.980  \\ 
  0.400 &   2.196 &   1.380 &   1.363 &   1.384 &   1.394 &   1.402 &   0.980 &   0.968 &   0.981  \\ 
  0.500 &   2.322 &   1.390 &   1.370 &   1.394 &   1.404 &   1.415 &   0.979 &   0.965 &   0.980  \\ 
  0.600 &   2.430 &   1.409 &   1.389 &   1.412 &   1.421 &   1.435 &   0.982 &   0.968 &   0.983  \\ 
  0.700 &   2.526 &   1.437 &   1.417 &   1.440 &   1.450 &   1.462 &   0.981 &   0.966 &   0.982  \\ 
  0.750 &   2.570 &   1.457 &   1.439 &   1.460 &   1.466 &   1.483 &   0.987 &   0.974 &   0.988  \\ 
  0.800 &   2.611 &   1.481 &   1.466 &   1.482 &   1.488 &   1.505 &   0.990 &   0.979 &   0.991  \\ 
  0.850 &   2.651 &   1.513 &   1.502 &   1.514 &   1.519 &   1.535 &   0.992 &   0.984 &   0.992  \\ 
  0.900 &   2.689 &   1.557 &   1.552 &   1.555 &   1.560 &   1.576 &   0.995 &   0.992 &   0.994  \\ 
  0.950 &   2.726 &   1.629 &   1.640 &   1.622 &   1.625 &   1.641 &   1.005 &   1.012 &   1.002  \\ 
  0.980 &   2.747 &   1.678 &   1.688 &   1.671 &   1.646 &   1.688 &   1.039 &   1.047 &   1.037  \\ 
\end{longtable}
}

\begin{figure}
\begin{center}
\includegraphics[width=0.85\columnwidth]{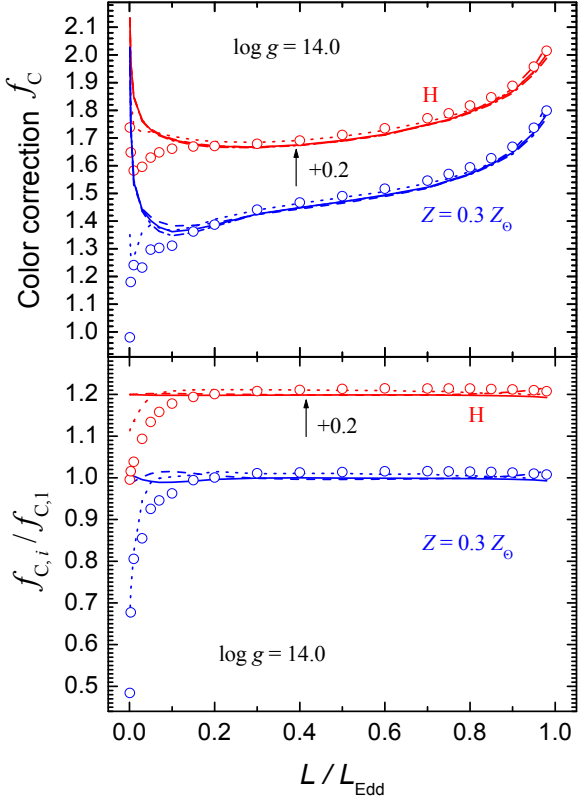}
\caption{\label{fig:fc}
{\it Top panel:} Dependence of the color correction factors  on the relative luminosity  for   the low surface
gravity ($\log g$ = 14.0) calculated by the five fitting procedures.
The results obtained with the first procedure are shown by the solid curve, 
the second procedure -- by the dashed curve, the third -- by the dot-dashed curve, the fourth --
by the dotted curve, and the fifth -- by circles. 
The lower curves are for the solar mixture of hydrogen and helium and $Z=0.3Z_{\odot}$. 
The upper curves correspond to pure hydrogen models and are shifted up by +0.2.
{\it Bottom panel:} Ratio of the color correction factors obtained using the second
(dashed curve), third (solid curve), fourth (dotted), and fifth (circles) procedures to the color correction factor from the first procedure. 
The curves corresponding to hydrogen models are shifted up by +0.2. 
 }
\end{center}
\end{figure}

\begin{figure}
\begin{center}
\includegraphics[width=0.85\columnwidth]{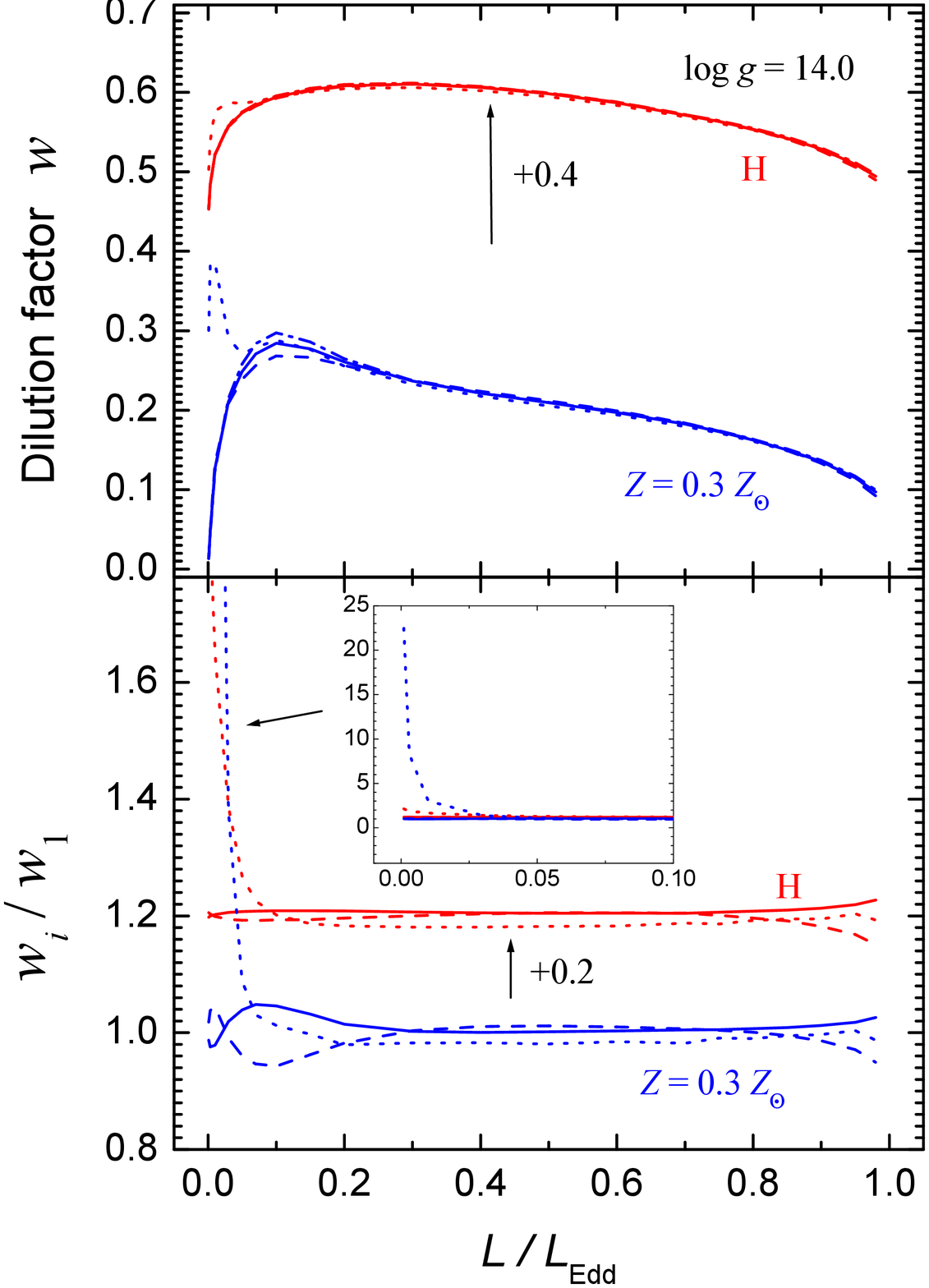}
\caption{\label{fig:wa}
Same as in Fig.\,\ref{fig:fc}, but for the dilution factor. The results are shown for the procedures 1--4. 
The curves corresponding to hydrogen models are shifted up for clarity. 
 }
\end{center}
\end{figure}

The obtained results are illustrated in Figs.\,\ref{fig:fc_main}--\ref{fig:wa}. 
The color-correction factors for models with various surface
gravities and chemical compositions versus relative luminosity are shown in Fig.\,\ref{fig:fc_main}.
A major feature for all dependences  $\fcol$--$L / L_{\rm Edd}$ is a local minimum of
$\fcol$ at some intermediate ($l \sim$ 0.1--0.5)  luminosity. The
color-correction factors are largest at luminosities closest to  the
Eddington luminosities  and decrease as the luminosity decreases. The  reason is
well known, it is due to a decreasing role of the Compton scattering in
comparison with true opacity (see Section \ref{sec:model}). At low
luminosity, the color-correction  factor increases again (see Fig. \ref{fig:fc_log}). 
At these conditions, Compton scattering is not so significant \citep{sw:07,Rauchetal:08}, 
and the increasing spectral color correction is related to the properties of the free-free 
opacity as first noted by \cite{Londonetal:84,Londonetal:86}. 
If the relative luminosity $l$  decreases below 0.2, the maximum
of the Planck function shifts to lower energies. At low photon energies, 
the free-free opacity dominates over electron scattering. In pure hydrogen or
helium atmospheres, radiation escapes at higher energies easier due to the strong, $\propto E^{-3}$, energy
dependence of the free-free opacity. This effect grows as $T_{\rm eff}$
decreases, and therefore, the spectrum becomes relatively harder. In
models including heavy elements, numerous absorption edges distort this picture, but
the common opacity properties are the same, the opacity increases as the  
photon energy decreases.

The computed dependences for model atmospheres with  heavy elements show relatively deep minima  in 
$\fcol$ at $l \sim 0.1$ (effective temperatures about 1 keV). 
These minima arise due to iron ions, because at those temperatures iron is not completely 
ionized  and a significant absorption edge of Fe {\sc xxv} and {\sc xxvi}  appears  at $\sim$ 9 keV
(Fig.~\ref{fig:fe}). 
 Due to this discontinuity in the spectra,  the color correction factors decrease. 
The depth of the minimum depends on the metal abundance. 
In the case of  solar abundance ($Z$ = $Z_\odot$) the minimum is
deep and it disappears in the case of $Z$ = 0.01 $Z_\odot$. At the same time the color
correction factor at high  luminosities depends very little on  $Z$.
At these luminosities, $\fcol$ depends mainly on a relative hydrogen fraction $X$,
being smaller for smaller $X$, i.e.  pure helium models have the smallest color correction factors.
The dependence of $\fcol$ on surface gravity is also insignificant at high
luminosities (see top panel of Fig.\,\ref{fig:fc_main}).

Examples of theoretical spectra fitted by the first and the fourth (one-parameter) 
procedures are shown in Fig.\,\ref{fig:fits}. For high luminosity models, 
both fits are identical, but at lower luminosities the difference is obvious. 
The  fourth procedure gives a better representation
of the overall spectral energy distributions, but the two-parameters fits (method 1) 
represent more accurately the theoretical spectra in the considered energy band. 
The difference between these two fits is more significant for the
model with heavy elements due to numerous absorption edges.

Residuals between theoretical spectra and the two fits, obtained using the first
and the second procedures versus photon energy  for
models with three relative luminosities ($l$ = 0.1, 0.5 and 0.95) and two
different chemical compositions ($X$=1 and $X$= 0.7374, solar composition) are
shown in Fig.\,\ref{fig:resid}. At photon energies $>$3--4 keV the fits have an
accuracy better than 2--3 \% for luminous models with heavy elements and for
all pure hydrogen models. The residuals for these models increase at energies
corresponding to the Wien tail ($E \ge$ 10 keV) due to exponential flux
decrease. Here the fits overestimate the computed spectra. On the contrary,
the blackbody fits underestimate the theoretical spectra at energies $<$2--3 keV
(see also Fig.\, \ref{fig:fits}) and the residuals grow here, too. The fits to the
spectrum of the low luminosity model with heavy elements are worse due to the
strong absorption edge at $\approx$ 9 keV. Fitting accuracy of pure helium
model atmosphere spectra is similar to pure hydrogen model spectra.
We can conclude that at high luminosities ($l \ge$ 0.5) the diluted blackbody spectra are 
rather good representations of the theoretical NS atmosphere model spectra within the {\it RXTE}/PCA energy band,
with  deviations not exceeding 2\%.

Absolute and relative comparisons of color corrections  $\fcol$ and
corresponding dilution factors $w_{\rm c}$ which were obtained using all four procedures
are shown in Figs.\,\ref{fig:fc} and \ref{fig:wa}. The 2-parameter
fitting procedures 1--3 give almost identical results differing  in $\fcol$ by not more than 2\%. 
The 1-parameter procedures 4--5 give also very similar $\fcol$ in the luminosity range  $l \ge 0.2$.   
The differences between 1- and 2-parameter  procedures become significant at lower luminosities, 
because these spectra  strongly deviate from  a diluted blackbody (these
deviations are significantly larger for models with metals).  
The fifth procedure employed by \citet{Madej.etal:04} and \citet{Madej:05} gives values of $\fcol$ 
at $l<0.2$ significantly lower than procedures 1--3.

\section{Comparison with previous calculations}
\label{sec:compar}

There is an extensive literature on NS atmospheres. The first detailed 
models are presented by \citet{Londonetal:84}. They considered 
hydrogen-helium-iron atmospheres with the relative number density ratio 
He/H=0.1/1 and various iron abundances (Fe/H=3.4$\times$10$^{-5}$=solar, 0.1 solar and 10$^{-5}$ of solar)
and accounted for Compton scattering using the Fokker-Planck approximation (Kompaneets equation). 
They showed that $\fcol$ (computed using 1-parameter procedure 4) is large at both high and low  ends of relative luminosities. 
If the metal abundance is low, $\fcol$ shows a broad minimum at $l\approx 0.2$ of about 1.4, 
while the models with solar Fe abundance also show a dip in $\fcol$ at $l\approx0.1$. 
The presence of the dip is also discussed in detail by \citet{Lapidusetal:86}. 
These results are consistent with our calculations at qualitative level. 

At luminosities close to Eddington, \citet{Pavlov.etal:91} derived a simple approximate formula for the ratio of
 surface temperature (which is not far from the color temperature) to $T_{\rm eff}$  
 \be
\frac{ T_{\rm surf}}{T_{\rm eff}} = \left( 0.14 \ln \frac{3+5X}{1-l} + 0.59\right) ^{-4/5} \left(  \frac{3+5X}{1-l} \right) ^{2/15} l ^{3/20} . 
 \ee
For our luminous models presented on Fig.\,\ref{fig:hhes} ($l=0.95$, $\log g=14.0$) this equation gives
$T_{\rm surf}/T_{\rm eff}$ = 1.518 ($X$=0), 1.569 ($X$=0.7374), and 1.582
($X$=1). In our models we have 1.534, 1.609 and 1.641, respectively. 
The relative deviations are 1.1, 2.5 and 3.7\%.

 We also compared our results for low temperature ($T_{\rm eff}$=1--3 MK, when Compton scattering is insignificant), pure H and pure He atmosphere models  to previous calculations by \citet{Zavlinetal:96} and found a good agreement \citep{sw:07}. 
Comparison of our LTE models with heavy elements to the non-LTE models computed using the {\sl TMAP} code \citep{Werner.etal:03} 
has been performed by \citet{Rauchetal:08}. 
It confirmed the reduction of the iron absorption edge at 9 keV for models with $T_{\rm eff} \approx 1$ keV due to iron over-ionization, 
as was found before by \citet{Londonetal:86}. However, the non-LTE effects reduce  the edge strength by one third only, 
much less than was claimed by \citet{Londonetal:86}, who considered only one level for  Fe {\sc xxvi}.

Recently, \citet{Madej:05} have computed a set of atmosphere
models for a small set of effective temperatures and  various $\log g=12.9,...,15.0$. 
They considered hydrogen-helium-iron atmospheres with the relative number density ratios 
Fe/He/H=3.7$\times$10$^{-5}$/0.11/1 corresponding to the `old' solar abundances with X=0.693.
They assumed LTE, ignored pressure ionization, 
and employed the integral formulation of Compton scattering using isotropic redistribution function.  
The radiative transfer equation was solved simultaneously with radiative equilibrium equation by the  iterative complete linearization method.
For  $\log = 14.6$ ($T_{\rm Edd}=2.42$ keV for $M=1.4M_{\odot}$) they obtained $\fcol=1.33, 1.26,1.40,1.54$ 
for $T_{\rm eff}=(1,1.5,2,2.5)\times10^7$ K which correspond to relative luminosities 
$l=0.016, 0.081, 0.26, 0.63$.
Our calculations (applying the fifth fit procedure used by \citealt{Madej:05})  
for the same $\log g$, $l$, and for $X=0.74, Z=Z_{\sun}$ (see Set 6 in Table 1 of the Online Material) 
give $f_{\rm c,5}=1.19, 1.27, 1.39, 1.52$. We see that at high luminosities, the difference is about 1\%, while
at the lowest temperature our color correction is significantly smaller. This can be explained
by the difference in the chemical abundance, since we have included among other elements also 
Ne, Mg, Al,  and S, which produce edges in the 1--3 keV energy range strongly affecting $\fcol$. 
Note that our pure hydrogen atmosphere model which lacks the edges gives $f_{\rm c,5}=1.39$ for $l=0.016$.
In general, we find the agreement rather satisfactory.

We remark, however, that the color-corrections obtained from the fifth procedure 
should not be compared to the data, because the color temperatures of the time-resolved spectra from 
X-ray bursts are computed by {\it fitting} the actual data in a specific energy interval (e.g. 3--20 keV) 
with the diluted blackbody function with arbitrary normalization. 
Thus, it is more appropriate to use color-correction factors obtained by  
procedures 1--3 (which give almost identical results).

It is also necessary to notice that obtaining $\fcol$ for low relative luminosity and low gravity (e.g. $\log g=14.0$) 
from the results presented in Table 1 of \citet{Madej:05} (that have minimum $l=0.06$) is impossible.  
One cannot scale the results obtained for $\log g=15.0$, because the opacity due to free-free and bound-free
transitions  depends on  temperature and  density in a different way, resulting for the same $l$
in smaller ionization and stronger absorption edges at higher $\log g$ (see Fig. \ref{fig:g}), affecting 
thus the color temperature estimation.

\section{Application to observational data}
\label{sec:appl}
 
\subsection{Method}

We discuss now how our theoretical calculations can be used for comparison with the observational data.
Observed spectra of X-ray bursting NSs are usually fitted by the blackbody \citep[see e.g.][]{GMH08}  
with two free parameters: a color temperature $T_{\rm bb}$ and a normalization $K=(R_{\rm bb}\ \mbox{[km]}/D_{10})^2$ (here 
$R_{\rm bb}$ is the apparent blackbody radius measured in km and $D_{10}$ is a distance to the source in units of 10 kpc). 
A total observed flux $F= R_{\rm bb}^2 \sigma_{\rm SB}T_{\rm bb}^4/D^2$ can be also expressed using basic NS parameters
$F=R^2\sigma_{\rm SB} T_{\rm eff}^4 (1+z)^{-2}/D^2$ \citep{lewin93}.  
Using the obvious relation 
$T_{\rm bb}=\fcol T_{\rm eff} (1+z)^{-1}$ we get
\be \label{eq:rinf}
R_{\infty}= R(1+z)  = R_{\rm bb} \fcol^2,  
\ee
where $R_{\infty}$ is the apparent NS radius at infinity. 
We can rewrite Eq. (\ref{eq:rinf}) as 
\be \label{u_fc}
      \fcol  = K^{-1/4}\, A,
\ee 
where 
\be\label{eq:defA}
A= \left( \frac{R_{\infty}\ \mbox{[km]}}{D_{\rm 10}} \right)^{-1/2} = \left( \frac{R\ \mbox{[km]}\ (1+z)}{D_{\rm 10}} \right)^{-1/2}
\ee
is a constant. 

\begin{figure}
\begin{center}
\includegraphics[width=1.0\columnwidth]{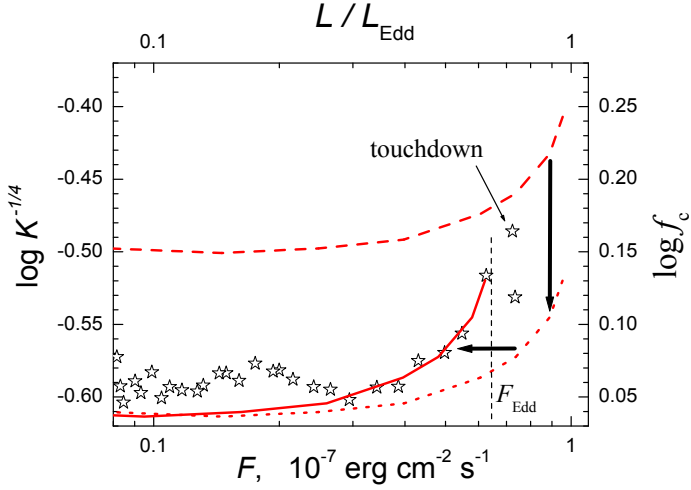}
\caption{\label{fig:data} 
The dependence $K^{-1/4}$--$F$ as observed during the cooling track of the long burst from 
4U\,1724--307 on November 8, 1996 (stars). 
The theoretical $\fcol$--$l$ dependence is shown by the dashed curve (right and upper axes) 
and the best-fit relation (solid curve). 
 }
\end{center}
\end{figure}

\citet{Ebisuzaki:87} suggested to fit cooling tracks presented in the form  $F/F_{\rm Edd}$--$T_{\rm bb}$ by the theoretical relation
$l$--$\fcol T_{\rm eff}$. 
However, the flux--temperature dependence is dominated by the approximate $F\propto T_{\rm bb}^4$ relation 
and therefore it is more appealing to emphasize the deviations from the constant apparent radius.  
In addition, \citet{Ebisuzaki:87} fixed a priori the value of $F_{\rm Edd}$ to the flux 
at the touchdown point, which corresponds to minimum $R_{\rm bb}$ and maximum $T_{\rm bb}$ (and flux). 
Such a restriction is not supported theoretically 
as the maximum flux observed from PRE X-ray bursts 
can exceed the flux when the photosphere is at the stellar surface by a factor $1+z$ \citep{lewin93}. 
A small apparent radius at the touchdown does not necessarily mean that the photosphere actually 
coincides with the NS surface, because at luminosities very close to Eddington 
the  color correction can be rather large \citep{Pavlov.etal:91}.
Recently, \citet{2010ApJ...722...33S} also argued in favor of this interpretation on the basis 
of the inconsistencies between the observables in some PRE bursts.

Because the  evolution of  $K^{-1/4}$ at sub-Eddington luminosities 
(at late burst stages) reflects the evolution of the color correction factor \citep{Penninx89,vP90,SP10}, 
we propose here to fit the observed  dependence $K^{-1/4}$--$F$  
by the theoretical relations $\fcol$--$(l\equiv L/L_{\rm Edd}\equiv F/F_{\rm Edd})$.  
The two free parameters of the fit are $A$ and 
the (observed) Eddington flux 
\be \label{u_fedd}
F_{\rm Edd} = \frac{L_{\rm Edd}}{4\pi\,D^2}(1+z)^{-2} = \frac{GMc}{\sigma_{\rm e}\,D^2}(1+z)^{-1}
\ee
(see Fig.\,\ref{fig:data}). 
Because the evolution of $\fcol$ is strongest near the Eddington flux,  
the PRE  X-ray bursts  are most suitable for the analysis. 
We would like to emphasize that the Eddington flux should be obtained 
from the fit to evolution of $K^{-1/4}$ in a broad range of luminosities. 

\begin{figure}
\begin{center}
\includegraphics[width=0.9\columnwidth]{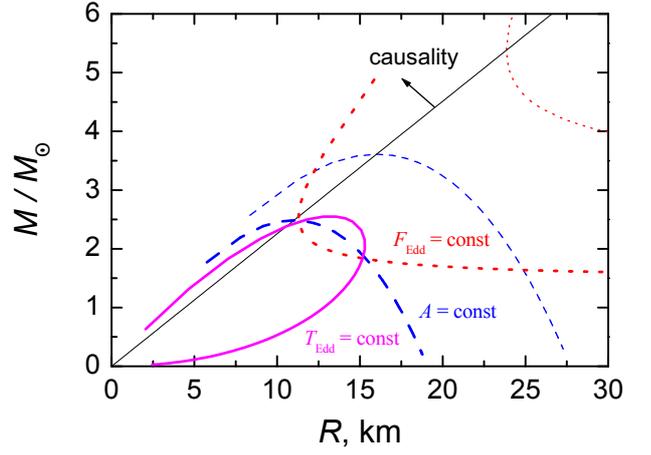}
\caption{\label{fig_curv} 
Constraints on $M$ and $R$ from various observables. 
The solid curve gives the relation obtained from the Eddington temperature (\ref{eq:tedd}) using Equations 
(\ref{eq:tedd2}) and (\ref{eq:m_r_u}).
Dotted curves are for the Eddington flux (\ref{eq:eddflux}), dashed curves are for $A$=const, shown  for 
two different distances. 
For the  distance  below the upper limit (\ref{eq:distmax}) there are solutions 
at the interception of the curves. If the assumed distance is too large, there are no solutions 
(the curves shown by thin lines do not cross). 
}
\end{center}
\end{figure}

Combining $A$ and $F_{\rm Edd}$ we can obtain the effective (Eddington) temperature corresponding to the Eddington flux 
on the NS surface corrected for the gravitational redshift: 
\be \label{eq:tedd}
T_{\rm Edd,\infty} =  \left( \frac{gc}{\sigma_{\rm SB} \sigma_{\rm e} } \right) ^{1/4} \frac{1}{1+z} = 6.4\times 10^9\ F_{\rm Edd}^{1/4} \ A^{-1}\ \mbox{K}.
\ee
This quantity is independent of the (uncertain) distance to the source and
can be used to express the NS radius through the observables and the compactness $u=R_{\rm S}/R=1-(1+z)^{-2}$: 
\be \label{eq:tedd2}
R =  \frac{c^3}{2  \sigma_{\rm e}  \sigma_{\rm SB} T_{\rm Edd,\infty}^4 } \ u\ (1-u)^{3/2} ,
\ee
and the mass is then found via (see solid curve in Fig.\,\ref{fig_curv})
\be \label{eq:m_r_u}
\frac{M}{M_{\odot}} =  \frac{R}{2.95\ {\rm km} } \ u .
\ee

If the distance $D$ is known (for example, if the source is in a globular cluster), one can get additional constraints on $M$ and $R$ 
(for a given chemical composition) from  $A$ and $F_{\rm Edd}$ separately. From the Eddington flux estimate we have (see dotted curves in Fig.\,\ref{fig_curv})
\bea \label{eq:eddflux}
R & =&  \frac {2 \sigma_{\rm e}  D^2 F_{\rm Edd} }{c^3 } \ u^{-1}\ (1-u)^{-1/2}  \nonumber \\
&=& 14.138\  {\rm km}\ (1+X)\ D_{10}^2\ F_{\rm Edd, -7}  \ u^{-1}\ (1-u)^{-1/2} ,  
\eea
where $F_{\rm Edd, -7}$ is the Eddington flux in units $10^{-7}$ erg cm$^{-2}$ s$^{-1}$ and 
 the mass is found using Equation (\ref{eq:m_r_u}). 
A measurement of $A$ gives another  constraint: 
\be
R = R_{\infty} \sqrt{1-u} = D_{10}\ A^{-2} \ \sqrt{1-u}\ \mbox{km} . 
\ee 
Combining with the parametric expression for the mass (\ref{eq:m_r_u}), we get the third 
relation between $M$ and $R$ shown by the dashed curves in Fig.\,\ref{fig_curv}. 

All three curves cross in one or two points (see Fig.\,\ref{fig_curv}) if the distance 
satisfies the following relation: 
\be \label{eq:distmax}
D_{10} \leq \frac{0.0177 }{(1+X)\ A^2\ F_{\rm Edd, -7} } . 
\ee
In the opposite case, there is no physical solution for $M$ and $R$ for given observables.

\subsection{Comparison to the data}

Although the analysis of the X-ray burst data is outside of the scope of the present paper, 
we give now a few examples of the observed bursts that can be interpreted using our approach. 
We also discuss a few cases where the data do not follow the theory and discuss the possible reasons for that. 

\citet{Penninx89} analyzed two long-duration ($>$100 s) PRE bursts observed by ME/{\it EXOSAT} in 1984 and 1986 
from 4U\,1608--52 in its hard state at a rather low persistent flux of (1--2)$\times 10^{-9}$ erg cm$^{-2}$ s$^{-1}$ in 2--20 keV band. 
The evolution of $T_{\rm bb} F^{-1/4}$ (which is proportional to $K^{-1/4}$) with $\log F$ shown in their Fig.\,7
is almost identical to our models presented in Fig.\,\ref{fig:fc_log}. 
Observations by LAC/{\it Ginga} of a long PRE burst from the atoll source 4U\,2129+11 during  its island (hard) state 
at a low persistent flux level of $\sim0.5\times 10^{-9}$ erg cm$^{-2}$ s$^{-1}$ are presented by \citet{vP90}.
The behavior of   $T_{\rm bb} F^{-1/4}$  at fluxes above 30\% of the peak (touchdown) flux 
shown in their Fig. 10 is very similar to that in our Fig. \ref{fig:fc_main}.
A long PRE burst was observed by {\it RXTE}\ on November 8, 1996 from 4U\,1724--307 in the hard state at a low accretion rate 
with persistent flux of $1.2\times 10^{-9}$ erg cm$^{-2}$ s$^{-1}$  \citep{MGL00,GMH08}. 
The evolution of $K^{-1/4}$ with flux during the cooling tail (shown in Fig. \ref{fig:data}, see also \citealt{SP10}) is almost identical to that of the burst from 4U\,2129+11. For both objects the data at high fluxes are well described  by the theory. We note that in both cases 
the position of the touchdown point in $K^{-1/4}$--$F$ diagram is not consistent with extrapolation of the data from intermediate fluxes, 
implying that the Eddington flux (at the NS surface) is actually smaller by about 15\% than the touchdown flux. This is consistent with the 
predictions that the maximum flux during a PRE burst (corresponding to the expanded photosphere) can be $1+z$ times larger than the surface Eddington flux \citep{lewin93}. 

However, not all bursts show an evolution consistent with theory. For example, two short PRE bursts 
from  4U\,1724--307 detected by {\it RXTE}\ on Feb 23 and May 22, 2004  during the high/soft state \citep{GMH08} 
show a nearly constant apparent area $K$ during the cooling stage \citep{SP10}. 
Four short PRE bursts from  4U\,1608--52
at a high persistent flux level of (3--6)$\times 10^{-9}$ erg cm$^{-2}$ s$^{-1}$ also demonstrate nearly constant area. 
This behavior is inconsistent with theoretical expectations. Therefore, we warn against using this kind of bursts 
for determining NS parameters  such as their masses and radii  \citep[cf.][]{GO10}. 

The most plausible reason for deviations of the data from theory is an influence of accretion  
on the emergent spectrum. The spectra and variability of the accreting NSs in the soft state are 
consistent with the presence of an optically thick boundary/spreading layer at the NS surface \citep{gil:03,rev:06}. 
The observed persistent spectra are well reproduced  in the framework of the spreading layer model \citep{IS99,Sul.Pout:06}. 
During the short bursts, accretion most probably persists and therefore   photons escape 
from the atmosphere of a rapidly  rotating spreading layer. 
The combined influence of the radiation pressure and centrifugal force cause a strong reduction of the effective gravity \citep{IS99} which 
causes the spectrum of the escaping radiation  to have a high value of $\fcol$ similar to that at $l\sim 1$ \citep{Sul.Pout:06}. 
This  effect may be responsible for nearly constant apparent areas in spite of the changes in the burst luminosity.

\section{Conclusions}
\label{sec:conclusions}

In this work we present an extended set of the X-ray bursting NS
model atmospheres in the luminosity range 0.001 -- 0.98 $L_{\rm Edd}$ for
three values of surface gravities ($\log g$ = 14.0, 14.3, and 14.6), and six
atmosphere chemical compositions (pure H and He, solar H/He mixture with $Z$ =
1, 0.3, 0.1 and 0.01 $Z_\odot$). Altogether 360 models were computed. The calculated
models are in a good agreement with models presented by other authors at
the same parameters. 

The spectra of  all models are redshifted and fitted  by diluted blackbody spectra in
the  {\it RXTE}/PCA energy band 3--20 keV. 
We use five different fitting procedures. At relatively large luminosities ($L \ge 0.3
L_{\rm Edd}$) all procedures give almost identical (difference less than 2 \%) 
color correction $\fcol = T_{\rm c}/T_{\rm eff}$. 
The color correction factor strongly depends on the relative luminosity and
the chemical composition and less significantly on the surface gravity.
For luminous models, color corrections depend mainly on the hydrogen mass
fraction $X$. Largest $\fcol$ are exhibited by pure hydrogen models ($X=1$).
The local minimum  in the $\fcol$--$L/L_{\rm Edd}$ dependence appears  
at about 10\% Eddington luminosity for models
with metals due to  an appreciable absorption edge at 9 keV arising from
 bound-free transitions in hydrogen-like iron. 
The depth of this minimum depends on the metal abundances and it disappears 
 in models with $Z$ = 0.01 $Z_\odot$.

The theoretical  emergent spectra and the color corrections with corresponding normalization 
are available online.\footnote{The spectra for all atmosphere models and the color correction factors 
can be found at the  T\"ubingen neutron star atmosphere model  web site http://astro.uni-tuebingen.de/$\sim$suleiman/web\_burst.}
These models can be used for interpretation of the X-ray bursting NS spectra.
We describe in detail the method for the evaluation  of NS masses and radii 
using the observed $K^{-1/4}$--$F$ dependence during cooling tails of PRE bursts. 
We also discuss the data that seem to follow the theory and also the bursts that contradict the theory. 
We argue that only the bursts that clearly follow theoretical models should be considered 
for further analysis with the aim to determine NS masses and radii.

\begin{acknowledgements}
VS was supported by DFG (grant SFB/Transregio 7 ``Gravitational Wave Astronomy'')
 and Russian Foundation for Basic Research (grant 09-02-97013 -p-povolzh'e-a).
JP has been supported by the Academy of Finland grant 127512.
\end{acknowledgements}


\end{document}